**SPINEX-Optimization: Similarity-based Predictions with Explainable Neighbors Exploration for Single, Multiple, and Many Objectives Optimization**

M.Z. Naser[1,2], Ahmed Z. Naser[3]
[1]School of Civil & Environmental Engineering and Earth Sciences (SCEEES), Clemson University, USA
[2]Artificial Intelligence Research Institute for Science and Engineering (AIRISE), Clemson University, USA
E-mail: mznaser@clemson.edu, Website: www.mznaser.com
[3]Department of Mechanical Engineering, University of Manitoba, Canada, E-mail: a.naser@umanitoba.ca

**Abstract**

This article introduces an expansion within SPINEX (Similarity-based Predictions with Explainable Neighbors Exploration) suite, now extended to single, multiple, and many objective optimization problems. The newly developed SPINEX-Optimization algorithm incorporates a nuanced approach to optimization in low and high dimensions by accounting for similarity across various solutions. We conducted extensive benchmarking tests comparing SPINEX-Optimization against ten single and eight multi/many optimization algorithms over 55 mathematical benchmarking functions and realistic scenarios. Then, we evaluated the performance of the proposed algorithm in terms of scalability and computational efficiency across low and high dimensions, number of objectives, and population sizes. The results indicate that SPINEX-Optimization consistently outperforms most algorithms and excels in managing complex scenarios – especially in high dimensions. The algorithm's capabilities in explainability, Pareto efficiency, and moderate complexity are highlighted through in-depth experiments and visualization methods.

**1.0 Introduction**

Optimization has been recognized as a cornerstone of scientific and engineering advancements to solve iterative to sophisticated real-world problems. One particular instance of a domain shift was the development of the simplex algorithm, which was then followed by the development of nonlinear programming techniques in the 1950s and 1960s, including the gradient descent method and its variants [1]. Then, the 1970s saw the rise of global optimization methods, such as simulated annealing and genetic algorithms, which could handle non-convex and discontinuous problems.

The field of optimization has expanded to include evolutionary and swarm algorithms. Evolutionary algorithms are those inspired by biological evolution. These include genetic algorithms, differential evolution, and evolution strategies. Such algorithms have gained prominence due to their ability to handle multi-modal and discontinuous objective functions. Then, swarm intelligence algorithms, such as particle swarm optimization and ant colony optimization, leverage collective behavior to explore solution spaces efficiently. Other nature-inspired methods, like simulated annealing and harmony search, have also demonstrated effectiveness in various optimization scenarios. Each of these methods offers a different mechanism of exploration and exploitation, balancing between intensively searching areas of the search space with known good solutions and exploring new regions [2].

As the significance of optimization algorithms spans various domains, such as engineering, economics, etc., to solve complex problems (i.e., resource allocation, scheduling, decision making, and system design etc.), the complexity of problems also grew [3]. Instead of handling single optimization problems, there is now a shift towards handling multiple and many-objective optimization problems. Thus, multi- and multiple-objective evolutionary algorithms (MOEAs) have been developed to find Pareto-optimal solutions that balance trade-offs between conflicting objectives. Popular MOEAs include NSGA-II, SPEA2, and MOEA/D, which employ different strategies to maintain diversity and convergence in the solution set [4,5].

Dealing with four or more objectives has presented new challenges [6]. For example, traditional Pareto-based approaches often struggle with the increased dimensionality of the objective space, leading to the development of new techniques. Decomposition-based, indicator-based, and reference point-based methods have emerged as promising approaches to tackle many-objective problems effectively. Parallel to the development of optimization algorithms, the importance of understanding and interpreting the optimization process has grown with the rise of explainable artificial intelligence (XAI). Such concepts are increasingly being incorporated into optimization frameworks to provide insights into the decision-making process and build trust in the solutions obtained [7]. This trend towards explainability is particularly crucial in high-stakes applications and domains (such as healthcare).

The use of similarity concepts in optimization as a means to overcome challenges about the existence of large number of objectives and poor insights into algorithmic decisions has gained traction as a means to improve the efficiency and effectiveness of optimization processes [8]. Similarity-based approaches leverage the concept that similar solutions in the decision space often exhibit similar performance in the objective space. In other words, such approaches leverage the notion that solutions 'close' to each other in the search space might yield similar performance levels. This assumption can be elemental in efficiently navigating the search space or scenarios with multiple or many objectives. By mapping the search space in terms of similarity, such algorithms can offer more directed and effective search strategies that reduce computational overhead while enhancing solution quality.

Similarity measures have been employed in various optimization contexts, including surrogate model-based optimization, to help select appropriate training samples and predict new solutions' performance [9]. Similarly, the same measures are also used in evolutionary algorithms to maintain population diversity and explore multiple optima simultaneously [10]. Moreover, similarity concepts have been utilized in local search procedures to balance exploitation and exploration [11]. In the context of many-objective optimization, where big data and large dimensionality pose significant challenges, similarity-based approaches can also offer a means to reduce the complexity of the search process by identifying and exploiting similarities among objectives. As such, these methods can potentially reduce the effective dimensionality of the problem and improve the scalability of optimization algorithms.

On a parallel front, integrating explainable exploration into optimization algorithms addresses can provide insights into why certain solutions are preferred over others, tracing the lineage of solution evolution and elucidating the influence of various problem features on the optimization outcomes. Here, we hypothesize that combining these two principles can be exploited to guide the search process, reduce computational costs, enhance the quality of obtained solutions, and challenge the traditional black-box optimization paradigms by fostering a more nuanced understanding of the solution landscape [12].

The above discussion elucidates the positives of integrating similarity-based predictions with explainable exploration for single, multiple, and many-objective optimization problems as a promising direction toward more efficient, scalable, and interpretable optimization techniques. Thus, this paper presents the development of a new algorithm, SPINEX (Similarity-based Predictions with Explainable Neighbors Exploration), that builds upon the concept of similarity and explainability for single, multiple, and many optimization problems. More specifically, we present this algorithm and its mechanisms; then, we examine the algorithm's performance on 55 diverse functions and scenarios compared to 18 commonly used algorithms.

## 2.0 Description of the SPINEX for optimization

*2.1 General description*
SPINEX-Optimization is an advanced, multi-faceted optimization technique designed to tackle single, multiple, and many objective problems across diverse domains. The algorithm's distinguishing feature is its use of multiple similarity metrics to quantify relationships between solutions. This multi-dimensional similarity assessment drives the algorithm's adaptive behavior and guides the search process. The algorithm calculates fitness values for all solutions in each iteration and constructs similarity matrices using the aforementioned metrics. These matrices are then leveraged to generate a transformation matrix, which is applied to the solution space to guide the search towards promising regions. This algorithm incorporates several mechanisms to enhance its adaptability. For example, the adaptive restart mechanism triggers a partial reinitialization of the population when progress stagnates to escape local optima. Then, the adaptive escape mechanism introduces controlled perturbations to maintain diversity and explore new areas of the solution space. The algorithm also features dimensional shifting to explore higher-dimensional spaces and potentially discover novel solutions temporarily. This adaptive nature extends to its parameters, with step sizes and population sizes dynamically adjusted based on the observed diversity and improvement rates. SPINEX maintains and evolves a Pareto front of non-dominated solutions for multi-objective optimization, allowing it to capture the trade-offs between competing objectives. The algorithm employs specialized techniques for updating and managing the Pareto front throughout the optimization process.

One of the key strengths of SPINEX is its emphasis on explainability as it incorporates mechanisms to track and analyze the influence of solutions on their neighbors, as well as the diversity within the population. This information is used not only to guide the optimization process but also to provide insights into the algorithm's decision-making and the characteristics of the solution space. To make use of such mechanisms, this algorithm incorporates a suite of visualization tools. These include plots of the fitness landscape, similarity matrices, convergence trajectories, and Pareto fronts for multi-objective problems. The algorithm also regularly generates detailed textual explanations of its progress and decision-making process.

*2.2 Detailed description*
A more detailed description of SPINEX's methods and functions is provided herein.

### Initialization (__init__)

The __init__ encapsulates the essential elements and sets up the operational parameters of SPINEX. The method signature is as follows:

```
def __init__(self, objective_function, n_variables, n_objectives=1, population_size=100, max_iterations=1000,
       similarity_methods=['correlation', 'cosine', 'spearman', 'euclidean'],
       allow_population_growth=False, is_multi_objective=False,
       verbose_explainability=False):
    self.objective_function = objective_function
    self.n_variables = n_variables
    self.n_objectives = n_objectives
    self.population_size = population_size
    self.max_iterations = max_iterations
    self.similarity_methods = similarity_methods
    self.iteration = 0
    self.weights = np.ones(len(similarity_methods)) / len(similarity_methods)
    self.best_solution = None
    self.best_fitness = float('inf')
    self.no_improvement_count = 0
    self.last_improvement_iteration = 0
    self.diversity_threshold = 1e-3
    self.allow_population_growth = allow_population_growth
    self.is_multi_objective = is_multi_objective
    self.step_size = 0.1
    self.step_size_min = 1e-6
    self.step_size_max = 1.0
    self.verbose_explainability = verbose_explainability
```

More specifically:
- **objective_function:** A callable that computes the fitness of a solution. It takes a solution as input and returns a fitness value (or values in the case of multi-objective optimization).
- **n_variables:** The number of decision variables in the optimization problem. This parameter determines the dimensionality of the solution space.
- **n_objectives (default = 1):** Specifies the number of objectives to optimize. A value greater than 1 indicates a multi-objective optimization problem.
- **population_size (default = 100):** The number of solution candidates in the population. This size can dynamically change if allow_population_growth is set to True.
- **max_iterations (default = 1000):** The maximum number of iterations (generations) to run the optimization process.
- **similarity_methods (default = ['correlation', 'cosine', 'spearman', 'euclidean']):** A list of similarity metrics used to compute similarities between solutions. These metrics guide the exploration and exploitation processes within the solution space.
- **allow_population_growth (default = False):** If set to True, the population size can increase during optimization to maintain diversity.
- **is_multi_objective (default = False):** A boolean indicating whether the optimization problem is multi-objective. This affects the fitness evaluation and the solution update strategy.
- **verbose_explainability (default = False):** When enabled, the algorithm provides detailed explanations of its operations at each iteration, aiding in understanding the decision-making process.
- **iteration:** Tracks the current iteration number within the optimization process.

- **weights:** Initializes equal weights for each similarity method. These weights may be adapted based on the performance contributions of each similarity metric throughout the optimization process.
- **best_solution and best_fitness:** Track the best solution found and its corresponding fitness. For multi-objective problems, best_fitness could be a vector.
- **no_improvement_count and last_improvement_iteration:** Monitor stagnation in the optimization process. no_improvement_count increments each time there is no improvement in the best fitness, and last_improvement_iteration records the iteration number of the last improvement.
- **diversity_threshold:** A threshold for measuring diversity in the population. If diversity falls below this value, mechanisms to reintroduce diversity are triggered.
- **step_size, step_size_min, and step_size_max:** Control the magnitude of changes applied to solutions during optimization. step_size adjusts adaptively based on the progress and diversity observed in the solution population.

## Method: calculate_similarity_matrices

This method initializes the solution space for the optimization algorithm and generates a population of solutions where each solution is a vector of n_variables dimensions. The values of these variables are randomly sampled from a uniform distribution over [0, 1], using np.random.rand(self.population_size, self.n_variables). This approach ensures a diverse starting point for the optimization process.

## Method: initialize_solution_space

This method computes similarity matrices based on the current population, represented by matrix X of shape (n_samples, n_features). This method handles various similarity measures to inform the optimization process and starts by normalizing the data:
- **Cosine Similarity:**
    - Computes the cosine of the angle between two vectors.
    - Equation: $similarity = \frac{X \cdot X^T}{\|X\|\|X^T\|}$
- **Correlation Similarity:**
    - Calculates the Pearson correlation coefficient.
    - Equation: $similarity = \text{corrcoef}(X)$
- **Euclidean Similarity:**
    - Inverted Euclidean distance, transforming the actual distances into a similarity measure.
    - Equation: $similarity = \frac{1}{1 + euclidean\_distance\ (X)}$
- **Spearman Similarity:**
    - Calculates the Spearman rank correlation between the columns of X, useful for capturing monotonic relationships between segments that may not necessarily be linear.
    - Equation: $similarity = \text{spearmanr}(X^T)[0]$
- **Combination of Similarity Matrices:**
    - Padding: Ensure all matrices have the same dimensions by padding smaller matrices.
    - Averaging: Calculate the average of these matrices to obtain a combined similarity matrix.
    - Equation: $S_{combined} = \frac{\sum_{i=1}^{N} S_i}{N}$, where $S_i$ is a similarity matrix for the i-th method, and N is the number of methods.

## Method: update_similarity_matrix

This function updates the similarity matrix when changes occur in the solution space, ensuring that the matrix reflects the most recent relationships between solutions.
- **Change Detection:**
    - Identify indices in the solution space where changes have occurred.

- **Matrix Update:**
  - For each similarity method, recalculate the similarity for changed parts of the solution space.
  - Efficiency Optimization: Instead of recalculating the entire matrix, only update the parts of the matrix corresponding to changed solutions.
- **Conditional Update:**
  - Use conditions to ensure matrix integrity and handle edge cases (e.g., scalar or one-dimensional outputs from similarity calculations).
  - Equation: $new\_matric[i,j] = similarity\_update(new\_space[i], new\_space[j])$

## Methodcalculate_transformation_matrix

This method develops a matrix that represents the relationships between different dimensions (variables) of the solutions based on their similarities. This matrix transforms the solution space to encourage exploration in directions indicated by the observed relationships.

- **Identity Matrix Initialization:**
  - Start with an identity matrix transformation of size (n_variables, n_variables). The identity matrix ensures that the base transformation is neutral, affecting no change if no cross-similarities influence it.
- **Cross-Similarity Calculation:**
  - Iterate over all pairs of different variables (i, j), i.e., where i does not equal j.
  - For each pair, compute the cross-similarity as the mean of the element-wise products of the i-th and j-th columns of the similarities matrix. This represents how changes in one variable correlate with changes in another across all solutions:
  - Assign this value to both the (i, j) and (j, i) positions in the matrix to ensure symmetry.
- **Matrix Symmetrization and Normalization:**
  - Symmetrize the matrix by averaging it with its transpose to guarantee that the transformation matrix is symmetric.
  - Add an identity matrix to preserve a degree of the original characteristics of the variables.
  - Normalize the transformation matrix by its Frobenius norm to ensure the transformed solutions remain scaled appropriately.

## Method: apply_transformation

This method function modifies the solution space using the transformation matrix calculated from the similarities between solution components. It also introduces controlled randomness to aid exploration.

- **Matrix Multiplication:**
  - Apply the transformation matrix to the solution space through matrix multiplication. This operation linearly combines the variables of each solution according to the transformation matrix, shifting the solution space toward regions suggested by the similarity-driven transformation.
- **Noise Introduction:**
  - Compute the noise intensity based on the current iteration and the step size, decreasing exponentially with the number of iterations. This decay helps reduce the scale of exploration as the algorithm converges, focusing more on exploitation:
  - Generate Gaussian noise with mean zero and standard deviation equal to the calculated noise intensity. The noise is added to the transformed solution space to introduce stochasticity, helping to escape local minima and explore new areas.
  - The final transformed solution space is the sum of the noise and the linearly transformed solutions.

## Method: intensify_search_around_best

This method focuses on searching for the best-performing solutions in the current population and is designed to exploit the promising regions of the solution space effectively.
- Selection of Best Solutions:
    - Sort the fitness values and select the top 10% of solutions. This is achieved by slicing the sorted indices of the fitness values.
    - Equation: $best\_indices = np.argsort(fitness\_values)[:0.1 \times self.population\_size]$
- Perturbation:
    - Apply a Gaussian perturbation with a standard deviation equal to the current step_size to these best solutions. This step aims to explore the local neighborhood of each selected solution, potentially uncovering better solutions near the current best ones.
    - Equation: $Perturbation = N(0, self.step\_size)$
    - Return the perturbed solutions, which will be evaluated in subsequent iterations.

## Method: extract_segments

This method provides a mechanism to reset the search process when there has been no improvement for a dynamically calculated number of iterations. Thus, this method helps in escaping deep local minima and reintroduces diversity into the population:
- Improvement Check:
    - Calculate the improvement rate since the last noted improvement and determine a threshold for restarting:
    - Equation = $improvement\_rate = (self.iteration - self.last\_improvement\_iteration)/self.max\_iterations$
- Restart Condition:
    - If the no_improvement_count exceeds the restart_threshold, reset the solution space by keeping the top 10% of the solutions and reintroducing new random solutions for diversity.
    - Equation: $new\_solutions[diversity\_indices] = np.random.rand(len(diversity\_indices), self.n\_variables)$
- Pareto Handling (for multi-objective):
    - If the problem is multi-objective, maintain the Pareto front among the retained solutions.

## Method: adaptive_escape_mechanism

This method dynamically adjusts the intensity of perturbations based on the current state of solution diversity and the duration of no improvement.
- Diversity and Severity Assessment:
    - Calculate the diversity of fitness values. For multi-objective optimization, use the average standard deviation across all objectives.
    - Determine the perturbation severity based on the duration of no improvement, capped at a maximum level:
    - Equation = $severity = min(self.no_improvement_count//10, 3)$
- Perturbation Application:
    - Applying a scaled Gaussian perturbation to randomly selected solutions depends on the calculated diversity and severity. Equation: $Perturbation = N(0, self.step\_size)$
    - Equation: $Perturbation = N(0, perturbation\_intensity)$

## Method: adaptive_parameters

This method dynamically adjusts the parameters of the optimization process based on the current state of the population, primarily focusing on step size adjustments to balance exploration and exploitation based on observed diversity.
- Diversity Calculation:

- For single-objective optimization, diversity is calculated as the standard deviation of the fitness values.
- For multi-objective optimization, diversity is the average of the standard deviations of each objective across the population.
- **Step Size Adjustment:**
  - If the diversity is below a predefined threshold (diversity_threshold), the step size is increased to encourage exploration. Conversely, if diversity is adequate, the step size is decreased to enhance exploitation.
  - The step size is further adjusted based on the no_improvement_count, promoting more exploration if there has been no improvement for more than 10 iterations.
  - The new step size is always clamped between step_size_min and step_size_max to maintain reasonable changes.
- **Verbose Output:**
  - If verbose_explainability is enabled, the method prints the current diversity and step size, providing insights into the algorithm's dynamic adjustments.

### Method: local_search

This function enhances optimization by performing a local search around a promising solution without requiring gradient information.

### Method: normalize_fitness

This function normalizes the fitness values within the range [0, 1] to facilitate selection processes or comparisons that benefit from a common scale.

- **Segment Extraction:** Both the target segment and a reference segment (usually the most recent one) are extracted.
- **Similarity Calculation:** The method calculates similarity scores using all available similarity methods, providing a detailed breakdown of how each method perceives the similarity.
- **Feature Contributions:** It calculates the absolute differences between the corresponding features of the two segments to determine which features contribute most to any dissimilarity.

### Method: adjust_population_size

This method adjusts the population size based on the diversity of the current population and the history of improvements. This allows the algorithm to adapt the number of solutions it maintains based on the problem's needs.

- **Diversity Assessment:**
  - Similar to the adaptive_parameters method, it uses the standard deviation of the fitness values to assess diversity.

- **Population Adjustment:**
  - If diversity is too low and population growth is allowed, increase the population size by 50% (up to a maximum of 1000) to introduce more variability.
  - Conversely, if there has been a significant lack of improvement, reduce the population size by 20% (down to a minimum of 50) to concentrate the search on the most promising solutions.

- **Reinitialization:**
  - If the population size is adjusted, reinitialize the solution space to the new population size, ensuring that the algorithm operates with an appropriately sized pool of solutions.

### Method: dimensional_shift

This method explores higher-dimensional transformations of the solution space to potentially discover better regions of the search space.

- **Conditional Dimension Expansion:**
    - With a 10% probability, the method augments the solution space by adding a new dimension filled with random values. This randomness can help escape local optima by introducing new characteristics to the solutions:
- **Covariance Matrix and Principal Components:**
    - Calculate the covariance matrix of the expanded solution space to understand the variance and correlation across all dimensions.
    - Perform eigen decomposition on the covariance matrix to extract the eigenvalues and eigenvectors, representing the data variance's principal components.
    - Select the top eigenvectors corresponding to the largest eigenvalues to form a new basis for the data's most significant directions.

- **Projection:** Project the expanded solution space onto these top eigenvectors. This operation effectively reduces the dimensionality back to the original while retaining the most informative directions of variation introduced by the new dimension.

### Method: track_solution_trajectory

This method creates a detailed record of how solutions transform over iterations. For each pair of original and transformed solutions, it records the original solution, the transformed solution, and the vector difference (shift vector) between them.

### Method: get_pareto_front

A utility method to determine the Pareto front from a given set of solutions and their fitness values.

### Method: update_pareto_front

This method integrates new solutions into an existing Pareto front for multi-objective optimization to ensure that the front represents the current set of non-dominated solutions. This also combines the current Pareto front with the new solutions and reevaluates the combined set's fitness.

### Method: optimize

This method orchestrates the overall optimization process by integrating various strategies and mechanisms to effectively explore and exploit the solution space, with special considerations for both single-objective and multi-objective optimization scenarios.

- **Initialization:**
    - The solution space is initialized with random values within [0, 1] for each variable.
    - Initialize variables to track the best solutions, Pareto front, trajectories, and explainability data.
- **Optimization Loop:**
    - Stagnation and Diversity Check: Regularly check for lack of variability in the solution space and reinitialize if necessary to avoid premature convergence.
    - Fitness Evaluation: Apply the objective function across the solution space to evaluate the fitness of each solution.
    - Similarity Calculation: Compute similarity matrices based on the current solution space to inform transformation strategies.
    - Explainability Features (if enabled): Generate explanations for actions taken during the optimization process, such as transformations and neighbor influences.

- **Solution Update:**
  - Best Solution Tracking: Continuously update the best found solution and check for convergence based on a tolerance threshold.
  - Pareto Front Update (for multi-objective optimization): Maintain and update the Pareto front as new non-dominated solutions are identified.
  - Transformation and Search Intensification: Apply transformation matrices and dimensional shifts to explore significant variations. Intensify the search around promising solutions by applying slight perturbations.
  - Adaptive Restart and Escape Mechanisms: Reinitialize or modify the solution space dynamically based on the lack of improvement over iterations and current solution diversity.
  - Population Adjustment: Optionally adjust the population size based on the diversity and search progress to maintain or increase the solution pool's robustness.
- **Termination:**
  - Convergence Check: If the optimization reaches the predefined tolerance for the objective function, terminate the optimization early.
  - Iteration Limit: Stop the optimization process after a predetermined number of iterations.
- **Final Outputs:**
  - For single-objective optimization, return the best solution found, its fitness, and the trajectory of solutions over iterations.
  - For multi-objective optimization, return the Pareto front, fitness values of the solutions on the front, and the trajectory.
- **Key Considerations**
  - Diversity and Adaptation: The method includes checks for constant solution columns and reinitializes the solution space if needed. This ensures that the algorithm does not get stuck in suboptimal regions of the solution space due to a lack of diversity.
  - Explainability and Debugging: Verbose outputs and detailed explainability data help understand the algorithm's behavior and diagnose issues during the optimization process.
  - Dynamic Parameter Adjustment: Adaptive mechanisms adjust search parameters like step size and population size in response to the current state of the optimization process, aiding in balancing exploration and exploitation effectively.

## Method: adjust_weights

This method dynamically adjusts the weights assigned to different similarity metrics based on their perceived effectiveness in improving the solutions. It aims to prioritize more effective methods by allocating greater weights to them.

- **Improvement Calculation:**
  - Compute the improvement ratio for each solution compared to the best fitness found so far. If best_fitness is zero (an optimal scenario), consider all improvements neutral to avoid division by zero: $improvements = fitness\_values/best\_fitness$
- **Contribution Estimation:**
  - Calculate the mean improvement contributed by each similarity method. This mean serves as a base to determine how much each method has contributed to finding better solutions.
- **Weight Update:**
  - Update the weights based on the negative exponential of the method contributions. This transformation ensures that methods leading to smaller improvements (thus larger contribution values) receive smaller weights: $new\_weights = exp(-method\_contributions)$
  - Normalize the weights so that they sum up to one, ensuring a proper probabilistic distribution.
- **Blending Old and New Weights**:
  - Blend the newly calculated weights with the old weights using a factor of 10% to introduce changes smoothly and maintain some stability in weight adjustments.

Method: identify_neighbors

This method, based on similarity scores, identifies the closest neighbors for each solution in the solution space. For each solution, sort the similarity scores and select the top solutions similar to neighbors, excluding the solution itself from its list of neighbors.

Method: analyze_neighbor_influence

This method quantifies the influence of neighboring solutions on each solution's fitness, helping to understand local interactions and dependencies. For each solution, the method calculates the difference in fitness between it and its neighbors. A positive difference indicates that the solution is better than its neighbors, while a negative difference suggests room for improvement.

Method: calculate_neighbor_diversity

This method evaluates the diversity of solutions in the neighborhood of each solution, providing insights into the local search space diversity. For example, this method calculates the diversity score for each solution based on the similarity scores with its neighbors. Diversity is computed as $1 - mean(neighbor\_similarities)$. A higher diversity score indicates that the solution has a more diverse set of neighbors, suggesting a more exploratory local search space.

Method: explain_iteration

This method provides detailed insights into the interactions and dynamics of the current solution space at a specific iteration by identifying neighbors based on similarity scores and calculating the influence and diversity relative to these neighbors.

*Additional functions:*

Method: fast_rankdata

This function computes the rank of each element in an array x, similar to scipy.stats.rankdata, but optimized for use with Numba. The algorithm works as follows:
- **Sorting Index**: The np.argsort(x) function is called to get the indices that would sort the array. This index array is stored in sorter.
- **Rank Assignment**: An array ranks of the same shape as sorter is created to store the ranks. Each position in ranks is then filled by mapping the sorted indices to their rank, which is simply their index in the sorted array plus one (to make ranks start from 1 instead of 0).
- **Mathematically**, this can be expressed as:
  - $rank[i] = position\ of\ x[i]\ in\ sorted\ x + 1$

Method: fast_spearman

This function computes the Spearman rank correlation coefficient for a matrix X, where rows represent samples and columns represent variables. The steps are.

Method: fast_euclidean

This function calculates the Euclidean distance between points represented in a normalized form, optimized for pairwise distance calculations between high-dimensional data points.

Method: visualize_explainability

This function leverages the data collected and processed during optimization to provide visual feedback on various aspects of the algorithm's performance and behavior.

The complete class of SPINEX is shown below:

```python
@jit(nopython=True)
def fast_rankdata(x):
    sorter = np.argsort(x)
    ranks = np.empty_like(sorter, dtype=np.float64)
    ranks[sorter] = np.arange(len(x)) + 1
    return ranks

@jit(nopython=True)
def fast_spearman(X):
    n, m = X.shape
    ranks = np.empty_like(X)
    for j in range(m):
        ranks[:, j] = fast_rankdata(X[:, j])
    return np.corrcoef(ranks.T)

@jit(nopython=True)
def fast_euclidean(X_norm, X_outer):
    sq_dists = X_norm[:, None]**2 + X_norm[None, :]**2 - 2*X_outer
    return 1 / (1 + np.sqrt(np.maximum(sq_dists, 1e-8)))

class SPINEX_Optimization:
    def __init__(self, objective_function, n_variables, n_objectives=1,
            population_size=100, max_iterations=1000,
            similarity_methods=['correlation', 'cosine', 'spearman', 'euclidean'],
            allow_population_growth=False, is_multi_objective=False,
            verbose_explainability=False):
        self.objective_function = objective_function
        self.n_variables = n_variables
        self.n_objectives = n_objectives
        self.population_size = population_size
        self.max_iterations = max_iterations
        self.similarity_methods = similarity_methods
        self.iteration = 0
        self.weights = np.ones(len(similarity_methods)) / len(similarity_methods)
        self.best_solution = None
        self.best_fitness = float('inf')
        self.no_improvement_count = 0
        self.last_improvement_iteration = 0
        self.diversity_threshold = 1e-3
        self.allow_population_growth = allow_population_growth
        self.is_multi_objective = is_multi_objective
        self.step_size = 0.1
        self.step_size_min = 1e-6
        self.step_size_max = 1.0
        self.verbose_explainability = verbose_explainability

    def initialize_solution_space(self):
```

```python
        return np.random.rand(self.population_size, self.n_variables)

    def calculate_similarity_matrices(self, X):
        n_samples, n_features = X.shape
        if n_samples < 2 or n_features < 2:
            return np.eye(n_samples)
        X_normalized = (X - X.mean(axis=0)) / (X.std(axis=0) + 1e-8)
        X_outer = np.dot(X, X.T)
        X_norm = np.linalg.norm(X, axis=1)

        def compute_similarity(method):
            try:
                if method == 'correlation':
                    return np.corrcoef(X.T)
                elif method == 'cosine':
                    norm_product = X_norm[:, None] * X_norm[None, :]
                    matrix = np.zeros_like(X_outer)
                    non_zero = norm_product != 0
                    matrix[non_zero] = X_outer[non_zero] / norm_product[non_zero]
                    return matrix
                elif method == 'spearman':
                    return fast_spearman(X)
                elif method == 'euclidean':
                    return fast_euclidean(X_norm, X_outer)
            except Exception as e:
                print(f"Error calculating {method} similarity: {e}")
                return None
        with ThreadPoolExecutor() as executor:
            similarity_matrices = list(executor.map(compute_similarity, self.similarity_methods))
        similarity_matrices = [mat for mat in similarity_matrices if mat is not None]
        if not similarity_matrices:
            return np.eye(n_samples)
        max_shape = max(mat.shape[0] for mat in similarity_matrices)
        padded_matrices = [np.pad(mat, ((0, max_shape - mat.shape[0]), (0, max_shape - mat.shape[1])),
                    mode='constant', constant_values=0) for mat in similarity_matrices]
        combined_similarity = np.mean(padded_matrices, axis=0)
        return np.nan_to_num(combined_similarity, nan=0.0, posinf=1.0, neginf=-1.0)

    def update_similarity_matrix(self, old_matrix, old_space, new_space):
        changed_indices = np.where(np.any(old_space != new_space, axis=1))[0]
        if len(changed_indices) == 0:
            return old_matrix
        new_matrix = old_matrix.copy()
        n_samples = new_space.shape[0]
        for method in self.similarity_methods:
            if method == 'correlation':
                update_matrix = np.corrcoef(new_space)
            elif method == 'cosine':
                update_matrix = 1 - cdist(new_space, new_space, metric='cosine')
```

```python
        elif method == 'spearman':
            update_matrix, _ = spearmanr(new_space)
        elif method == 'euclidean':
            distances = cdist(new_space, new_space, metric='euclidean')
            update_matrix = 1 / (1 + distances)
        if np.isscalar(update_matrix):
            update_matrix = np.array([[1.0]])
        elif update_matrix.ndim == 1:
            update_matrix = update_matrix.reshape(1, 1)
        if update_matrix.shape != (n_samples, n_samples):
            padded_matrix = np.full((n_samples, n_samples), np.nan)
            padded_matrix[:update_matrix.shape[0], :update_matrix.shape[1]] = update_matrix
            update_matrix = padded_matrix
        new_matrix[changed_indices] = update_matrix[changed_indices]
        new_matrix[:, changed_indices] = update_matrix[:, changed_indices]
    return new_matrix

def calculate_transformation_matrix(self, similarities):
    transformation = np.eye(self.n_variables)
    for i in range(self.n_variables):
        for j in range(self.n_variables):
            if i != j:
                cross_similarity = np.mean(similarities[:, i] * similarities[:, j])
                transformation[i, j] = cross_similarity
    transformation = (transformation + transformation.T) / 2
    transformation += np.eye(self.n_variables)
    return transformation / np.linalg.norm(transformation)

def apply_transformation(self, solution_space, transformation):
    transformed_space = np.dot(solution_space, transformation)
    noise_intensity = self.step_size * np.exp(-0.05 * self.iteration)
    noise = np.random.normal(0, noise_intensity, solution_space.shape)
    return transformed_space + noise

def intensify_search_around_best(self, solution_space, fitness_values):
    best_indices = np.argsort(fitness_values)[:int(0.1 * self.population_size)]
    best_solutions = solution_space[best_indices]
    perturbation = np.random.normal(0, self.step_size, best_solutions.shape)
    return best_solutions + perturbation

def adaptive_restart_mechanism(self, solution_space, fitness_values):
    improvement_rate = (self.iteration - self.last_improvement_iteration) / self.max_iterations
    restart_threshold = 30 * (1 + improvement_rate)
    if self.no_improvement_count > restart_threshold:
        print(f"Restarting at iteration {self.iteration} due to no improvement...")
        self.no_improvement_count = 0
        self.last_improvement_iteration = self.iteration
        n_keep = int(0.1 * self.population_size)
        if not self.is_multi_objective:
```

```python
            best_indices = np.argsort(fitness_values.flatten())[:n_keep]
        else:
            pareto_front = self.get_pareto_front((solution_space, fitness_values))
            best_indices = [np.where((solution_space == sol).all(axis=1))[0][0] for sol in pareto_front[:n_keep]]
        new_solutions = self.initialize_solution_space()
        new_solutions[:len(best_indices)] = solution_space[best_indices]
        diversity_indices = np.random.choice(range(len(best_indices), self.population_size),
                                             size=int(0.2 * self.population_size), replace=False)
        new_solutions[diversity_indices] = np.random.rand(len(diversity_indices), self.n_variables)
        return new_solutions
    return solution_space

def adaptive_escape_mechanism(self, solution_space, fitness_values):
    if not self.is_multi_objective:
        diversity = np.std(fitness_values)
    else:
        diversity = np.mean([np.std(fitness_values[:, i]) for i in range(fitness_values.shape[1])])
    severity = min(self.no_improvement_count // 10, 3)
    perturbation_intensities = [0.1, 0.3, 0.6, 1.0]
    perturbation_intensity = perturbation_intensities[severity] * self.step_size
    if diversity < self.diversity_threshold:
        perturbation_intensity *= 2.0
    perturb_mask = np.random.choice([0, 1], size=solution_space.shape[0], p=[0.5, 0.5])
    perturbation = np.random.normal(0, perturbation_intensity, solution_space.shape)
    solution_space[perturb_mask == 1] += perturbation[perturb_mask == 1]
    return np.clip(solution_space, 0, 1)

def adaptive_parameters(self, fitness_values):
    if not self.is_multi_objective:
        diversity = np.std(fitness_values)
    else:
        diversity = np.mean([np.std(fitness_values[:, i]) for i in range(fitness_values.shape[1])])
    if diversity < self.diversity_threshold:
        self.step_size *= 1.1
    else:
        self.step_size *= 0.9
    self.step_size = np.clip(self.step_size, self.step_size_min, self.step_size_max)
    if self.no_improvement_count > 10:
        self.step_size *= 0.9
    else:
        self.step_size *= 1.1
    self.step_size = np.clip(self.step_size, self.step_size_min, self.step_size_max)
    if self.verbose_explainability:
        print(f"Diversity: {diversity:.4f}")
        print(f"Updated step size: {self.step_size:.4f}")

def local_search(self, solution):
    result = minimize(self.objective_function, solution, method='Nelder-Mead', options={'maxiter': 100})
    return result.x
```

```python
def normalize_fitness(self, fitness_values):
    min_fitness = np.min(fitness_values)
    max_fitness = np.max(fitness_values)
    if np.isclose(max_fitness, min_fitness):
        return np.ones_like(fitness_values)
    return (fitness_values - min_fitness) / (max_fitness - min_fitness)

def adjust_population_size(self, fitness_values):
    diversity = np.std(fitness_values)
    old_size = self.population_size
    if diversity < self.diversity_threshold:
        if self.allow_population_growth:
            self.population_size = min(int(1.5 * self.population_size), 1000)
    elif self.no_improvement_count > 20:
        self.population_size = max(int(0.8 * self.population_size), 50)
    if self.population_size != old_size:
        print(f"Adjusting population size from {old_size} to {self.population_size}")
    return self.initialize_solution_space()[:self.population_size]

def dimensional_shift(self, solution_space):
    if np.random.rand() < 0.1:
        expanded_space = np.column_stack([solution_space, np.random.rand(self.population_size, 1)])
        cov_matrix = np.cov(expanded_space, rowvar=False)
        eigenvalues, eigenvectors = np.linalg.eigh(cov_matrix)
        top_eigenvectors = eigenvectors[:, np.argsort(eigenvalues)[::-1][:self.n_variables]]
        return np.dot(expanded_space, top_eigenvectors)
    return solution_space

def track_solution_trajectory(self, original_space, transformed_space):
    return [{
        'original': orig,
        'transformed': trans,
        'shift_vector': trans - orig
    } for orig, trans in zip(original_space, transformed_space)]

def update_pareto_front(self, current_front, new_solutions):
    if len(current_front) == 0:
        return list(new_solutions)
    combined = np.vstack((current_front, new_solutions))
    combined_fitness = np.array([self.objective_function(sol) for sol in combined])
    return list(self.get_pareto_front((combined, combined_fitness)))

@staticmethod
def get_pareto_front(solutions_and_fitness):
    solutions, fitness_values = solutions_and_fitness
    if len(solutions) == 0:
        return []
    is_pareto = np.ones(len(solutions), dtype=bool)
```

```python
        for i, fitness_i in enumerate(fitness_values):
            if is_pareto[i]:
                dominated = np.all(fitness_values <= fitness_i, axis=1) & np.any(fitness_values < fitness_i, axis=1)
                is_pareto[dominated] = False
                is_pareto[i] = True
        return list(solutions[is_pareto])

    def optimize(self, tolerance=1e-6):
        solution_space = self.initialize_solution_space()
        trajectory = []
        pareto_front = [] if self.is_multi_objective else None
        pareto_fitness = [] if self.is_multi_objective else None
        self.best_fitness = float('inf') if not self.is_multi_objective else None
        self.best_solution = None
        self.explainability_data = []
        for self.iteration in range(self.max_iterations):
            try:
                if np.all(np.std(solution_space, axis=0) < 1e-8):
                    solution_space += np.random.normal(0, 1e-5, solution_space.shape)
                    solution_space = np.clip(solution_space, 0, 1)
                non_constant_cols = ~np.all(solution_space == solution_space[0, :], axis=0)
                if np.sum(non_constant_cols) < 2:
                    print("Not enough non-constant columns. Reinitializing solution space.")
                    solution_space = self.initialize_solution_space()
                    continue
                solution_space_filtered = solution_space[:, non_constant_cols]
                if solution_space_filtered.shape[1] == 0:
                    print("All columns are constant. Reinitializing solution space.")
                    solution_space = self.initialize_solution_space()
                    continue
                fitness_values = np.apply_along_axis(self.objective_function, 1, solution_space)
                similarities = self.calculate_similarity_matrices(solution_space)
                if self.verbose_explainability and self.iteration % 10 == 0:
                    neighbors = self.identify_neighbors(solution_space, similarities)
                    influence = self.analyze_neighbor_influence(solution_space, neighbors, fitness_values)
                    diversity = self.calculate_neighbor_diversity(similarities, neighbors)
                    explanation = self.explain_iteration(solution_space, similarities, fitness_values, self.iteration)
                    print(explanation)
                    min_length = min(len(influence), len(diversity), len(fitness_values), len(solution_space))
                    self.explainability_data.append({
                        'iteration': self.iteration,
                        'influence': influence[:min_length],
                        'diversity': diversity[:min_length],
                        'fitness_values': fitness_values[:min_length],
                        'similarities': similarities[:min_length, :min_length],
                        'solution_space': solution_space[:min_length],
                    })
                if not self.is_multi_objective:
                    current_best = np.argmin(fitness_values)
```

```python
                    if fitness_values[current_best] < self.best_fitness:
                        self.best_solution = solution_space[current_best].copy()
                        self.best_fitness = fitness_values[current_best]
                        self.best_solution = self.local_search(self.best_solution)
                        self.best_fitness = self.objective_function(self.best_solution)
                        if not np.isscalar(self.best_fitness):
                            self.best_fitness = self.best_fitness[0]
                        trajectory.append((self.best_solution.copy(), self.best_fitness))
                        self.no_improvement_count = 0
                        self.last_improvement_iteration = self.iteration
                        if np.isclose(self.best_fitness, 0, atol=tolerance):
                            print(f"Global optimum found at iteration {self.iteration}.")
                            break
                    else:
                        self.no_improvement_count += 1
                else:
                    old_pareto_size = len(pareto_front)
                    new_pareto_front = self.get_pareto_front((solution_space, fitness_values))
                    pareto_front = self.update_pareto_front(pareto_front, new_pareto_front)
                    pareto_fitness = [self.objective_function(sol) for sol in pareto_front]
                    if len(pareto_front) > 0:
                        trajectory.append((pareto_front[-1].copy(), pareto_fitness[-1]))
                    if len(pareto_front) > old_pareto_size:
                        self.no_improvement_count = 0
                    else:
                        self.no_improvement_count += 1
                if self.iteration % 100 == 0:
                    if not self.is_multi_objective:
                        print(f"Iteration {self.iteration}: Best Fitness = {self.best_fitness}")
                    else:
                        print(f"Iteration {self.iteration}: Pareto Front Size = {len(pareto_front)}")
                transformation_vector = np.mean(similarities, axis=0)
                transformation_matrix = np.diag(transformation_vector[:self.n_variables])
                original_space = solution_space.copy()
                solution_space = self.apply_transformation(solution_space, transformation_matrix)
                solution_space = self.dimensional_shift(solution_space)
                if not self.is_multi_objective:
                    best_solutions = self.intensify_search_around_best(solution_space, fitness_values)
                    solution_space[:len(best_solutions)] = best_solutions
                solution_space = self.adaptive_restart_mechanism(solution_space, fitness_values)
                solution_space = self.adaptive_escape_mechanism(solution_space, fitness_values)
                if self.allow_population_growth or self.population_size > len(solution_space):
                    solution_space = self.adjust_population_size(fitness_values)
                self.adaptive_parameters(fitness_values)
            except Exception as e:
                print(f"Error in iteration {self.iteration}: {e}")
                import traceback
                traceback.print_exc()
                break
```

```python
        if self.is_multi_objective:
            return pareto_front, pareto_fitness, trajectory
        else:
            return self.best_solution, self.best_fitness, trajectory

    def adjust_weights(self, fitness_values, best_fitness):
        epsilon = 1e-10
        if best_fitness == 0:
            improvements = np.ones_like(fitness_values)
        else:
            improvements = np.divide(fitness_values, best_fitness, out=np.ones_like(fitness_values), where=best_fitness!=0)
        method_contributions = np.zeros(len(self.similarity_methods))
        for idx in range(len(self.similarity_methods)):
            method_contributions[idx] = np.mean(improvements)
        new_weights = np.exp(-method_contributions)
        new_weights_sum = np.sum(new_weights)
        if new_weights_sum > epsilon:
            new_weights /= new_weights_sum
        else:
            new_weights = np.ones_like(new_weights) / len(new_weights)
        self.weights = 0.9 * self.weights + 0.1 * new_weights

    def identify_neighbors(self, solution_space, similarities, k=5):
        neighbors = []
        for i in range(solution_space.shape[0]):
            sim_scores = similarities[i]
            nearest_indices = np.argsort(sim_scores)[-k-1:-1][::-1]
            neighbors.append(nearest_indices)
        return neighbors

    def analyze_neighbor_influence(self, solution_space, neighbors, fitness_values):
        influence_scores = []
        for i, neighbor_indices in enumerate(neighbors):
            neighbor_fitness = fitness_values[neighbor_indices]
            fitness_diff = fitness_values[i] - neighbor_fitness
            influence = np.mean(fitness_diff)
            influence_scores.append(influence)
        return np.array(influence_scores)

    def calculate_neighbor_diversity(self, similarities, neighbors):
        diversity_scores = []
        for i, neighbor_indices in enumerate(neighbors):
            neighbor_similarities = similarities[i, neighbor_indices]
            diversity = 1 - np.mean(neighbor_similarities)
            diversity_scores.append(diversity)
        return np.array(diversity_scores)

    def explain_iteration(self, solution_space, similarities, fitness_values, iteration):
```

```python
            k = min(5, solution_space.shape[0] - 1)
            neighbor_indices = np.argsort(similarities, axis=1)[:, -k-1:-1]
            neighbor_fitness = np.take(fitness_values, neighbor_indices, axis=0)
            if fitness_values.ndim == 1:
                fitness_diff = fitness_values[:, np.newaxis] - neighbor_fitness
            else:
                fitness_diff = np.mean(fitness_values[:, np.newaxis, :] - neighbor_fitness, axis=2)
            influence = np.mean(fitness_diff, axis=1)
            neighbor_similarities = np.take_along_axis(similarities, neighbor_indices, axis=1)
            diversity = 1 - np.mean(neighbor_similarities, axis=1)
            explanation = f"\nIteration {iteration} Explanation:\n"
            explanation += f"Average neighbor influence: {np.mean(influence):.4f}\n"
            explanation += f"Average neighbor diversity: {np.mean(diversity):.4f}\n"
            for i in range(min(3, solution_space.shape[0])):
                explanation += f"\nSolution {i}:\n"
                if fitness_values.ndim == 1:
                    explanation += f"  Fitness: {fitness_values[i]:.4f}\n"
                else:
                    explanation += f"  Fitness: {fitness_values[i]}\n"
                explanation += f"  Neighbor influence: {influence[i]:.4f}\n"
                explanation += f"  Neighbor diversity: {diversity[i]:.4f}\n"
                explanation += f"  Top 3 similar neighbors: {neighbor_indices[i, -3:]}\n"
            return explanation

    def plot_enhanced_influence_diversity(self):
        if not self.explainability_data:
            print("No explainability data available.")
            return
        last_data = self.explainability_data[-1]
        solution_space = last_data['solution_space']
        fitness_values = last_data['fitness_values']
        influence = last_data['influence']
        diversity = last_data['diversity']
        min_length = min(len(influence), len(diversity), len(fitness_values))
        influence = influence[:min_length]
        diversity = diversity[:min_length]
        fitness_values = fitness_values[:min_length]
        if len(fitness_values.shape) > 1:
            fitness_for_color = np.sum(fitness_values, axis=1)
        else:
            fitness_for_color = fitness_values
        plt.figure(figsize=(12, 8))
        scatter = plt.scatter(influence, diversity, c=fitness_for_color, cmap='viridis', alpha=0.6)
        plt.colorbar(scatter, label='Fitness (sum for multi-objective)')
        if not self.is_multi_objective:
            best_index = np.argmin(fitness_values)
            plt.scatter(influence[best_index], diversity[best_index], color='red', s=300, marker='*', label='Best Solution')
            top_5_indices = np.argsort(fitness_values)[:5]
```

```python
            plt.scatter(influence[top_5_indices], diversity[top_5_indices], color='orange', s=75, marker='o', label='Top 5 Solutions')
            for i, idx in enumerate(top_5_indices):
                plt.annotate(f'S{i+1}', (influence[idx], diversity[idx]), xytext=(5, 5), textcoords='offset points')
        else:
            pareto_front = self.get_pareto_front((solution_space[:min_length], fitness_values))
            pareto_indices = [np.where((solution_space[:min_length] == sol).all(axis=1))[0][0] for sol in pareto_front]
            plt.scatter(influence[pareto_indices], diversity[pareto_indices], color='red', s=100, marker='*', label='Pareto Front')
            for i, idx in enumerate(pareto_indices):
                plt.annotate(f'P{i+1}', (influence[idx], diversity[idx]), xytext=(5, 5), textcoords='offset points')
        plt.xlabel('Neighbor Influence')
        plt.ylabel('Neighbor Diversity')
        plt.title('Neighbor Influence vs Diversity (Final Population)')
        plt.legend()
        plt.axvline(x=0, color='gray', linestyle='--', alpha=0.5)
        plt.axhline(y=np.mean(diversity), color='gray', linestyle='--', alpha=0.5)
        plt.text(0.95, 0.95, 'High Diversity\nHigh Influence', ha='right', va='top', transform=plt.gca().transAxes)
        plt.text(0.05, 0.95, 'High Diversity\nLow Influence', ha='left', va='top', transform=plt.gca().transAxes)
        plt.text(0.95, 0.05, 'Low Diversity\nHigh Influence', ha='right', va='bottom', transform=plt.gca().transAxes)
        plt.text(0.05, 0.05, 'Low Diversity\nLow Influence', ha='left', va='bottom', transform=plt.gca().transAxes)
        plt.show()

    def plot_fitness_landscape(self):
        if not self.explainability_data:
            print("No explainability data available. Make sure verbose_explainability is set to True.")
            return
        last_data = self.explainability_data[-1]
        solution_space = last_data['solution_space']
        fitness_values = last_data['fitness_values']
        if solution_space.shape[1] > 2:
            pca = PCA(n_components=2)
            solution_space_2d = pca.fit_transform(solution_space)
        else:
            solution_space_2d = solution_space
        plt.figure(figsize=(12, 8))
        if self.is_multi_objective:
            plt.scatter(fitness_values[:, 0], fitness_values[:, 1], c=solution_space_2d[:, 0], cmap='viridis')
            plt.xlabel('Objective 1')
            plt.ylabel('Objective 2')
            plt.title('Fitness Space (Multi-objective)')
            plt.colorbar(label='First PCA component')
        else:
            x = np.linspace(solution_space_2d[:, 0].min(), solution_space_2d[:, 0].max(), 100)
            y = np.linspace(solution_space_2d[:, 1].min(), solution_space_2d[:, 1].max(), 100)
            X, Y = np.meshgrid(x, y)
            from scipy.interpolate import griddata
            Z = griddata(solution_space_2d, fitness_values, (X, Y), method='cubic')
            ax = plt.subplot(111, projection='3d')
```

```python
        surf = ax.plot_surface(X, Y, Z, cmap='viridis', alpha=0.8)
        ax.scatter(solution_space_2d[:, 0], solution_space_2d[:, 1], fitness_values, c='r', marker='o')
        ax.set_xlabel('Dimension 1')
        ax.set_ylabel('Dimension 2')
        ax.set_zlabel('Fitness')
        ax.set_title('Fitness Landscape (Single-objective)')
        plt.colorbar(surf)
    plt.show()

def plot_similarity_matrix(self):
    if not self.explainability_data:
        print("No explainability data available. Make sure verbose_explainability is set to True.")
        return
    last_data = self.explainability_data[-1]
    similarities = last_data['similarities']
    plt.figure(figsize=(10, 8))
    sns.heatmap(similarities, cmap='viridis', annot=False)
    plt.title('Solution Similarity Matrix')
    plt.show()

def plot_convergence(self):
    if not hasattr(self, 'trajectory'):
        print("No trajectory data available.")
        return
    iterations, fitness_values = zip(*self.trajectory)
    plt.figure(figsize=(10, 6))
    plt.plot(iterations, fitness_values)
    plt.xlabel('Iteration')
    plt.ylabel('Best Fitness')
    plt.title('Convergence Plot')
    plt.yscale('log')
    plt.show()

def plot_pareto_front(self):
    if not self.is_multi_objective:
        print("This method is only applicable for multi-objective optimization.")
        return
    if not hasattr(self, 'pareto_fitness'):
        print("No Pareto front data available.")
        return
    pareto_fitness = np.array(self.pareto_fitness)
    plt.figure(figsize=(10, 6))
    plt.scatter(pareto_fitness[:, 0], pareto_fitness[:, 1])
    plt.xlabel('Objective 1')
    plt.ylabel('Objective 2')
    plt.title('Pareto Front')
    plt.show()

def plot_final_solution_and_neighbors(self):
```

```python
        if not self.explainability_data:
            print("No explainability data available.")
            return
        last_data = self.explainability_data[-1]
        solution_space = last_data['solution_space']
        fitness_values = last_data['fitness_values']
        if not self.is_multi_objective:
            best_index = np.argmin(fitness_values)
            best_solution = solution_space[best_index]
            k = 5
            distances = np.linalg.norm(solution_space - best_solution, axis=1)
            neighbor_indices = np.argsort(distances)[1:k+1]
            plt.figure(figsize=(10, 8))
            plt.scatter(solution_space[:, 0], solution_space[:, 1], c=fitness_values, cmap='viridis', alpha=0.6)
            plt.colorbar(label='Fitness')
            plt.scatter(best_solution[0], best_solution[1], c='r', s=200, marker='*', label='Best Solution')
            plt.scatter(solution_space[neighbor_indices, 0], solution_space[neighbor_indices, 1],
                        c='orange', s=100, marker='o', label='Nearest Neighbors')
            for i, idx in enumerate(neighbor_indices):
                plt.annotate(f'N{i+1}', (solution_space[idx, 0], solution_space[idx, 1]))
            plt.xlabel('Dimension 1')
            plt.ylabel('Dimension 2')
            plt.title('Best Solution and Its Neighbors')
            plt.legend()
            plt.show()

    def visualize_explainability(self):
        try:
            self.plot_enhanced_influence_diversity()
        except Exception as e:
            print(f"Error in plot_enhanced_influence_diversity: {e}")
        try:
            self.plot_final_solution_and_neighbors()
        except Exception as e:
            print(f"Error in final_solution_and_neighbors: {e}")
        try:
            self.plot_fitness_landscape()
        except Exception as e:
            print(f"Error in plot_fitness_landscape: {e}")
        try:
            self.plot_similarity_matrix()
        except Exception as e:
            print(f"Error in plot_similarity_matrix: {e}")
        try:
            self.plot_convergence()
        except Exception as e:
            print(f"Error in plot_convergence: {e}")
        if self.is_multi_objective:
            try:
```

```
        self.plot_pareto_front()
    except Exception as e:
        print(f"Error in plot_pareto_front: {e}")
```

## 3.0 Description of benchmarking algorithms

We examined the performance of SPINEX against a comprehensive set of optimization algorithms. The single-objective optimizers included Simulated Annealing, Differential Evolution, Basin-hopping, Nelder-Mead, Cuckoo Search, Grey Wolf Optimizer, Bat Algorithm, Harmony Search, Flower Pollination Algorithm, and Biogeography-Based Optimization. We also evaluated multi-objective and many objectives optimization algorithms such as NSGA-II, MOEA/D, R-NSGA-II, NSGA-III, SMS-EMOA, AGE-MOEA, UNSGA-III, and SPEA2. These algorithms are particularly adept at handling scenarios where multiple and many objectives need to be balanced, such as accuracy versus computational cost or model complexity. Each of these algorithms is described in this section (with additional details available from the cited sources). Tables 1 and 2 compare these algorithms.

*3.1 Algorithms used in single objective optimization experiments*

3.1.1 Simulated Annealing (SA)

The SA is a probabilistic technique for approximating the global optimum of a given function. This algorithm is inspired by the process of annealing in metallurgy, which involves heating and controlling the cooling of a material to increase the size of its crystals and reduce their defects [13]. The SA starts with a high temperature that allows it to explore the search space freely, accepting worse solutions with a certain probability of avoiding local minima. As the temperature lowers, the algorithm becomes less likely to accept worse solutions, increasingly focusing on exploring the neighborhood of the current best solution. This process continues until a stopping condition is met, typically a temperature close to zero or a maximum number of iterations. The SA can escape local minima. However, its main limitation is its sensitivity to the cooling schedule—the rate at which the temperature is decreased.

3.1.2 Differential Evolution (DE)

The DE algorithm was introduced by Storn and Price in 1997 [14] to optimize problems over continuous spaces. The DE works by initializing a population of potential solutions and iteratively improving them through mutation, crossover, and selection operations. The mutation in DE involves creating a trial vector by adding the weighted difference between two population vectors to a third vector. This helps in maintaining diversity within the population and prevents premature convergence. The crossover operation then mixes the trial vector with another population vector to increase the potential search space the population explores. Finally, selection involves choosing the better vector between the trial and target vectors for the next generation. As one can see, the DE has a simple mechanism and can efficiently handle non-differentiable, noisy, and time-varying objective functions. However, one of the main limitations of the DE includes a dependency on the fine-tuning of its control parameters, such as mutation factor and crossover rate, which can significantly affect its performance.

### 3.1.3 Basin-hopping (BH)

The Basin-hopping is a global optimization algorithm designed to escape the local minima traps and is also known as Iterated Local Search. This algorithm was popularized by Wales and Doye in 1997 [15] in molecular conformation studies. The algorithm combines a local minimization method with a stochastic hopping step that attempts to escape the basin of attraction of local minima. In practice, Basin-hopping modifies a randomly selected component of the solution vector, uses a local minimization algorithm to find the nearest local minimum, and then decides whether to accept this new point based on the Metropolis criterion, which allows occasional uphill moves to avoid local optima traps. The process iteratively repeats with the hope of "hopping" out of local minima and finding increasingly better solutions. The BH can find global minima in landscapes where local minima are densely packed and traditional gradient-based approaches fail. Some of its limitations include its reliance on the efficiency of the underlying local minimization technique and the need for careful calibration of the 'temperature' parameter used in the Metropolis criterion.

### 3.1.4 Nelder-Mead Algorithm (NMA)

The Nelder-Mead algorithm (aka. the simplex method) is a heuristic search method for unconstrained optimization that Nelder and Mead developed in 1965 [16]. The NMA does not require any derivative information, which makes it suitable for optimizing non-differentiable, discontinuous, and noisy functions. The NMA evaluates the function at hand at each vertex and applies operators such as reflection, expansion, contraction, and shrinkage to move the simplex towards lower values on the function surface. These operations adjust the simplex in response to the optimized function's landscape, seeking areas with lower function values. The NMA can handle complex optimization problems where derivatives are unavailable or difficult to compute. However, the NMA can be slow in convergence compared to gradient-based methods, especially near the optimum.

### 3.1.5 Cuckoo Search (CS)

The Cuckoo Search is an optimization algorithm inspired by the obligate brood parasitism of some cuckoo species by laying their eggs in the nests of other host birds. This algorithm was developed by Yang and Deb in 2009 [17]. The algorithm uses a small population of solutions and replaces less promising solutions with potentially better ones, which are generated randomly by leveraging the Lévy flight behavior—a random walk that is more efficient in exploring the search space due to its heavy-tailed step length distribution. This algorithm can balance exploration and exploitation through the adaptive use of Lévy flights for global exploration and local random walks for exploitation. This makes it particularly effective for solving nonlinear and multimodal global optimization problems. On the other hand, the dependency on fine-tuning its control parameters, such as the discovery rate of alien eggs and the step size distribution, can significantly affect its performance.

### 3.1.6 Grey Wolf Optimizer (GWO)

The Grey Wolf Optimizer is an optimization algorithm inspired by grey wolves' social hierarchy and hunting behavior, developed by Seyedali Mirjalili et al. in 2014 [18]. GWO simulates the leadership hierarchy among grey wolves, categorized into alpha, beta, delta, and omega wolves, where the alpha leads the hunt, followed by beta and delta, with omega at the bottom of the hierarchy. The strength of Grey Wolf Optimizer lies in its robustness and ability to deal with various optimization problems, including nonlinear, non-differentiable, and multimodal functions. The GWO can exhibit slow convergence rates in the latter stages of the search, particularly when approaching the global optimum.

3.1.7 Bat Algorithm (BA)
The Bat Algorithm was developed by Yang in 2010 [19] as an optimization method inspired by the echolocation behavior of bats. This algorithm models how bats use varying pulse rates of sound and loudness to navigate and hunt for prey. Each bat represents a solution in this algorithm, and it uses simulated sonar echoes to determine the distance to an optimum. Bats fly randomly at a velocity, adjusting their wavelength and loudness as they approach their target. The frequency of the sound pulses, which affects how finely a bat can search around its current location, starts high to encourage exploration and decreases as a bat home in on a target, promoting exploitation. The algorithm integrates these behaviors to balance local and global search effectively.

3.1.8 Harmony Search (HS)
The HS algorithm was developed by Geem et al. in 2001 [20]. This algorithm is inspired by the musical process of searching for a perfect state of harmony. Harmony Search simulates this process by using a "harmony memory," which holds several solution vectors similar to a set of musical instruments. New harmonies (solutions) are improvised in each iteration based on three rules: memory consideration, pitch adjustment, and random consideration. These rules mimic the musicians' playing by choosing tones from their memory, adjusting them slightly, or choosing new random tones, thereby exploring the solution space. The HS has high flexibility and effectiveness in dealing with discrete and continuous optimization problems but can exhibit slow convergence in highly competitive landscapes or in cases where the search space is excessively large.

3.1.9 Flower Pollination Algorithm (FPA)
This is another algorithm developed by Yang et al. in 2012 [21]. The FPA is inspired by the natural pollination process of flowering plants and uses the behavior of pollinators like insects and birds and the strategies of global and local pollination. FPA distinguishes between biotic and abiotic pollination, which involves pollinators that can travel long distances, promoting global search, and abiotic refers to local pollination caused by wind and other factors. In the algorithm, global pollination is simulated by Lévy flights, which efficiently explore the search space due to their ability to make large jumps, while local pollination is modeled by simple diffusion processes for intensification near the current best solutions. This dual mechanism helps in balancing the exploration and exploitation phases effectively.

3.1.10 Biogeography-Based Optimization (BBO)
The BBO algorithm was developed by Dan Simon in 2008 [22]. The BBO is inspired by biogeography – the study of the distribution of biological species in different geographical areas over historical periods. BBO models the mathematical optimization problem as the problem of species distribution, where each habitat corresponds to a possible solution, and the habitat suitability index (HSI) represents the solution's fitness. The algorithm employs migration and mutation mechanisms to share information between habitats, simulating species migration to enhance the solutions. Habitats with high HSI (high-quality solutions) have more emigration, while those with low HSI accept more immigrants. This process helps refine the solutions by exploring new possibilities and exploiting known good solutions. Some of the BBO's limitations include a dependency on migration rates and mutation probabilities and possible premature convergence if the diversity of solutions is not maintained throughout the search process.

Table 1 A comparison among the examined algorithms in this study

| Feature | SA | DE | BH | NMA | CS | GWO | BA | HS | FPA | BOO |
|---|---|---|---|---|---|---|---|---|---|---|
| Core Mechanism | Probabilistic minimization with temperature cooling | Mutation, crossover, and selection | Local minimization with stochastic steps | Simplex-based direct search | Lévy flights and host nest analogy | Social hierarchy and hunting mimicry | Echolocation behavior simulation | Musical improvisation process | Biotic and abiotic pollination simulation | Species migration and mutation |
| Primary Strength | Escapes local minima effectively | Robust in multimodal landscapes | Global optimum search in rugged landscapes | Derivative-free, handles non-smooth functions | Balances exploration and exploitation | Effective in diverse optimization tasks | Suitable for complex problems | Flexible and effective for discrete problems | Handles nonlinear and multimodal problems | Models complex problems efficiently |
| Main Limitation | Sensitive to cooling schedule | Parameter sensitive | Dependent on local minimization efficiency | Poor performance in high dimensions | Sensitive to control parameters | Risk of premature convergence | Dependent on parameter tuning | Parameter tuning required | Sensitive to control parameters | Needs careful calibration of rates |
| Typical Applications | General optimization, especially where derivatives are unavailable | Engineering, bioinformatics | Chemical structure optimization | Real-time systems, parameter estimation | Engineering design, scheduling | Renewable energy, robotics | Structural design, aerospace | Engineering optimization, water networks | Engineering optimization, environmental modeling | Electrical engineering, environmental modeling |
| Scalability | Moderate | High | Moderate | Low | Moderate | Moderate to high | Moderate | Moderate | Moderate | Moderate |

*3.2 ML algorithms used in multiple and many objectives optimization experiments*
3.2.1 Non-dominated Sorting Genetic Algorithm II (NSGA-II)
The NSGA-II was developed by Deb et al. in 2002 [23] as an evolutionary algorithm to solve multi-objective optimization problems and enhance its predecessor, NSGA. This version addresses issues such as computational complexity and the need to specify NSGA sharing parameters. NSGA-II introduces a fast non-dominated sorting approach combined with a crowding distance mechanism, which helps preserve the population's diversity across generations without requiring any additional parameters for niche preservation. The strength of NSGA-II lies in its robustness and ability to find a diverse set of solutions along the Pareto optimal front, and it is particularly effective in handling two to three-objective problems. However, NSGA-II tends to struggle with the number of objectives increases, particularly in terms of computational efficiency and maintaining diversity among solutions. The algorithm can also struggle with scalability issues when dealing with a very high-dimensional search space or many objectives.

3.2.2 Multi-Objective Evolutionary Algorithm based on Decomposition (MOEA/D)

The MOEA/D algorithm MOEA/D was developed by Zhang and Li in 2007 [5]. MOEA/D is fundamentally different from other evolutionary algorithms like NSGA-II since it decomposes a multi-objective optimization problem into a set of scalar optimization subproblems and solves them simultaneously using evolutionary techniques. Each subproblem is associated with a weight vector, and the aggregation of these weight vectors defines the shape and extent of the Pareto front that the algorithm attempts to approximate. The core of MOEA/D's methodology lies in its use of neighborhood-based selection and recombination, wherein each subproblem only exchanges information with its neighboring subproblems based on the proximity of their weight vectors. This localized interaction enhances the algorithm's efficiency and accelerates convergence towards the Pareto front. MOEA/D is robust in handling complex objective spaces and can effectively distribute computational resources over subproblems. This makes it particularly effective for problems with conflicting objectives or non-uniformly distributed across the Pareto front. However, the performance of MOEA/D heavily relies on the proper setting of decomposition strategies and the distribution of weight vectors. On the same front, this algorithm's efficiency can degrade if the number of objectives is very high or if the objectives are highly correlated.

3.2.3 Reference-point Based NSGA-II (R-NSGA-II)

The R-NSGA-II was introduced by Deb and Jain in 2014 as an innovative variant of the widely-used NSGA-II algorithm [24]. R-NSGA-II integrates the concept of reference points, which are user-specified targets in the objective space that guide the search toward desired regions of the Pareto front. The R-NSGA-II enhances the original NSGA-II by incorporating a novel selection mechanism that minimizes the distance from these reference points and the crowding distance and non-domination sorting used in NSGA-II. The R-NSGA-II can provide tailored solutions that align closely with the user's preferences and can efficiently converge to regions of the Pareto front that are most relevant to the decision makers. On the other hand, the R-NSGA-II's performance heavily depends on the placement and number of reference points provided by the user, which makes it less efficient in scenarios with a high number of objectives or when a broad coverage of the Pareto front is required.

3.2.4 Non-dominated Sorting Genetic III (NSGA-III)

The third variant of the NSGA algorithm is an extension of the aforementioned NSGA-II [25]. This variant is specifically designed to handle optimization problems with a large number of objectives. NSGA-III utilizes a set of predefined reference points that are systematically distributed across the objective space to guide the selection process and maintain a diverse set of solutions evenly spread across the Pareto front. The core mechanism of NSGA-III involves two main steps: non-dominated sorting and niche preservation. The former ensures that the solutions closer to the Pareto front are given priority, similar to NSGA-II, while the latter is handled differently using reference points. This method is particularly effective when dealing with many objectives because it prevents the dominance mechanism from overly concentrating on a few objectives and ignoring others. The performance of NSGA-III can be limited by the distribution and number of reference points used, and the setup of these points requires careful consideration. It can also significantly affect the algorithm's ability to capture all relevant regions of the Pareto front.

### 3.2.5 S-metric Selection Evolutionary Multi-Objective (SMS-EMOA)

The SMS-EMOA was developed by Emmerich [26] in 2005. This algorithm utilizes the hypervolume indicator (i.e., S-metric) to quantify the volume covered by the population in the objective space and offer a direct measure of convergence and diversity. The algorithm selects individuals that contribute most to the increase in hypervolume and progressively pushes the population towards the Pareto front while maintaining diverse solutions. SMS-EMOA. The distinctive feature of SMS-EMOA is its selection process: at each generation, it evaluates each individual's contribution to the overall hypervolume and eliminates the one with the smallest contribution, replacing it with new offspring. This method ensures that the population evolves towards regions of the objective space that are not well represented, effectively maximizing the covered volume. This makes the SMS-EMOA particularly effective when a well-balanced trade-off between convergence and diversity is crucial (aka. engineering design and multi-criteria decision-making processes). The algorithm can be computationally expensive as calculating the hypervolume contribution becomes increasingly challenging as the number of objectives grows.

### 3.2.6 Approximation-Guided Evolutionary Multi-Objective Evolutionary Algorithm (AGE-MOEA)

The AGE-MOEA adopts a unique approach using approximation techniques to guide the evolutionary process [27]. The core idea behind AGE-MOEA is to construct approximations of the Pareto front at each generation, which then inform the selection and generation of new solutions. This process helps to focus the evolutionary efforts on areas of the objective space that are poorly represented or closer to the theoretical Pareto front, thus improving both convergence rates and solution diversity. The AGE-MOEA evaluates each individual's fitness by its performance relative to other individuals and by how well it fits within the current approximation of the Pareto front. This dual approach drives the population towards more accurately representing the Pareto front over successive generations. The AGE-MOEA can rapidly converge to high-quality solutions as its approximations can significantly reduce the number of generations needed to achieve satisfactory solutions. However, AGE-MOEA's limitations arise when dealing with highly complex or deceptive problem landscapes, as the approximation quality significantly affects the algorithm's performance, and poor approximations can lead to premature convergence or neglect of significant regions of the objective space.

### 3.2.7 UNSGA-III (Unified NSGA-III)

This advanced variant of NSGA-III addresses various challenges encountered in many-objective optimization problems [28]. The UNSGA-III builds on the robust framework of NSGA-III by enhancing its capabilities in diversity maintenance and convergence in scenarios involving a high number of objectives. The UNSGA-III introduces modifications that improve its ability to handle constraints and adapt to different problem complexities. The algorithm initially sorts the population into fronts based on non-dominance criteria. It then employs a novel selection mechanism that considers the proximity to reference points, as in NSGA-III, and integrates additional diversity and convergence measures. This ensures that the selected individuals contribute positively to achieving a well-rounded representation of the Pareto front. It is worth noting that the limitations of UNSGA-III mirror those of NSGA-III, particularly in terms of the computational effort required to manage and process a large set of reference points and the potential for decreased efficiency in cases where the reference points are not optimally distributed.

3.2.8 Strength Pareto Evolutionary Algorithm 2 (SPEA2)

SPEA2 was developed by Zitzler et al. in 2001 [29] as an enhancement of the original SPEA and is designed to improve the efficiency and effectiveness of multi-objective optimization. This algorithm is notable for its incorporation of a fine-grained fitness assignment strategy and a density estimation technique, which aids in preserving diversity among solutions while pushing toward the Pareto front. SPEA2 operates with a unique archive where a secondary population where elite solutions are stored to maintain historical information about the best solutions found. The main loop of SPEA2 involves generating offspring from the archived solutions through crossover and mutation, then selecting the next generation based on strength (dominance count) and density (distance to nearest neighbors). These measures ensure that individuals who dominate more solutions or are located in less crowded regions have higher survival chances, promoting a well-distributed set of Pareto-optimal solutions. The SPEA2 has a robust mechanism for balancing the exploration and exploitation aspects of the search process. However, SPEA2's limitations primarily stem from its archive management and density estimation processes, as the archive size needs to be carefully managed to avoid excessive growth. Additionally, while effective in promoting diversity, the density estimation technique can be computationally intensive, especially as the number of objectives increases.

Table 2 A comparison among the examined algorithms in this study

| Feature | NSGA-II | MOEA/D | R-NSGA-II | NSGA-III | SMS-EMOA | AGE-MOEA | UNSGA-III | SPEA2 |
|---|---|---|---|---|---|---|---|---|
| Core Mechanism | Fast non-dominated sorting, crowding distance | Decomposition into scalar subproblems | Reference point based selection | Reference point based approach for many-objectives | Hypervolume-based selection | Approximation-guided selection | Enhanced NSGA-III with unified handling of complexities | Strength and density-based selection, external archive |
| Primary Strength | Efficient on 2-3 objectives, maintains diversity | Efficient distribution of solutions, adaptable via decomposition methods | Targets specific regions via reference points | Scales well to many-objective problems, maintains diversity | Direct focus on improving hypervolume, robust in diverse applications | Rapid convergence, effective in scenarios with limited computational resources | Enhanced adaptability and constraint management | Effective management of historical solutions, balanced exploration and exploitation |
| Main Limitation | Struggles with many-objectives, high computational cost with large populations | Performance depends on the decomposition strategy, struggles with non-separable objectives | Dependent on the choice and placement of reference points | Requires careful selection of reference points, high computational effort | Computationally expensive with increasing objectives, reference point dependent | Dependent on the quality of approximations, can lead to premature convergence | Similar to NSGA-III, complexity in tuning | Archive management can be cumbersome, computationally intensive density estimation |
| Scalability | Moderate | High | Moderate | High | Low | Moderate | High | Moderate |
| Diversity Preservation | High | Moderate | High | High | High | Moderate | High | High |

## 4.0 Description of benchmarking experiments, metrics, and datasets

This benchmarking analysis involves a set of 45 mathematical benchmarking functions and 10 realistic scenarios. This analysis was run and evaluated in a Python 3.10.5 environment using an Intel(R) Core(TM) i7-9700F CPU @ 3.00GHz and a RAM of 32.0GB. All algorithms were run in default settings to allow fairness and ensure reproducibility. For example, the algorithms were run for a collection of population sizes (50, 100, 200), dimensions (2, 5, 10, 25, and 50), and iterations (100, 1000).

### 4.1 Mathematical benchmarking functions

Thirty single objective mathematical benchmarking functions were used in the first component of this examination. These functions span various landscapes and complexity levels and are presented in Table 3 (also see Fig. 1). The selected functions are commonly used in optimization benchmarks, as noted in [30,31]. The goal of such optimization testing is to minimize these functions at two dimensions.

Table 3 List of mathematical benchmarking functions

| Function Name | Type | Characteristics | Challenges |
|---|---|---|---|
| Ackley | Multimodal | Exponential term causes flat regions | Difficult to converge due to flat regions |
| Beale | Multimodal | Steep valleys and flat regions | Difficult to locate the global minimum |
| Bent cigar | Unimodal | Long, narrow, bent ridge to global minimum | Navigating the narrow, bent ridge |
| Booth | Unimodal | Smooth, simple, convex | Too straightforward, not challenging |
| Branin | Multimodal | Known for its distinct local minima | Identifying the correct local minimum |
| Bukin | Multimodal | Steep drops, narrow global minimum region | Difficulty due to steepness and narrowness |
| Bukin N6 | Multimodal | Steep drop-offs and narrow global minimum | Navigating drop-offs is challenging |
| Cosine mixture | Multimodal | A combination of flat areas and steep drops | Flat areas make convergence difficult |
| Cross-in-tray | Multimodal | Deep holes and cross-structure | Optimizing around deep, narrow holes |
| DeJong | Unimodal | Simple, convex | Overly simplistic for most modern methods |
| Dixon-Price | Unimodal | Steep, quadratic-like valley | Steepness complicates optimization |
| Drop-wave | Multimodal | Sharp peaks and deep holes | Extreme gradients challenge optimization |
| Eggholder | Multimodal | Non-convex with large holes | Handling non-convexity and large variations |
| Griewank | Multimodal | Oscillatory | Managing product term, many local optima |
| Himmelblau | Multimodal | Multiple global minima | Multiple solutions can confuse optimization |
| Levy | Multimodal | Complex with many local minima | Sharp peaks make optimization tricky |
| Matyas | Unimodal | Smooth, convex | Simplicity limits its utility |
| Perm 0, d, beta | Multimodal | Dependence on constants, d, and beta | Customization required varies widely |
| Rastrigin | Multimodal | Highly oscillatory | Locating global minimum amid numerous local minima |
| Rosenbrock | Unimodal | Non-convex, banana-shaped valley | Sensitive to initial conditions, slow convergence |
| Salomon | Multimodal | Rugged landscape with sharp peaks | Landscape makes gradient-based methods struggle |
| Schaffer N2 | Multimodal | Simple structure with sharp peaks | The sharpness of peaks presents difficulty |
| Schaffer N4 | Multimodal | Complex structure with sharp peaks/valleys | Highly challenging due to its sharp variations |
| Schwefel | Multimodal | Irregular, distant global minimum | Remote and hard-to-find global minimum |
| Six-hump camel | Multimodal | Two valleys, six humps | Complex landscape with multiple optima |
| Sphere | Unimodal | Simple, smooth, convex | Simple |
| Step | Unimodal | Discontinuous steps create flat plateaus | Discontinuities make gradient use impossible |
| Three-hump camel | Multimodal | Has three humps, as the name suggests | Multiple optima make it complex |
| Trid | Unimodal | Linear terms create a diagonal ridge | Unconventional landscape complicates optimization |
| Zakharov | Unimodal | Smooth, convex near global minimum | Deceptive simplicity masks optimization difficulty |

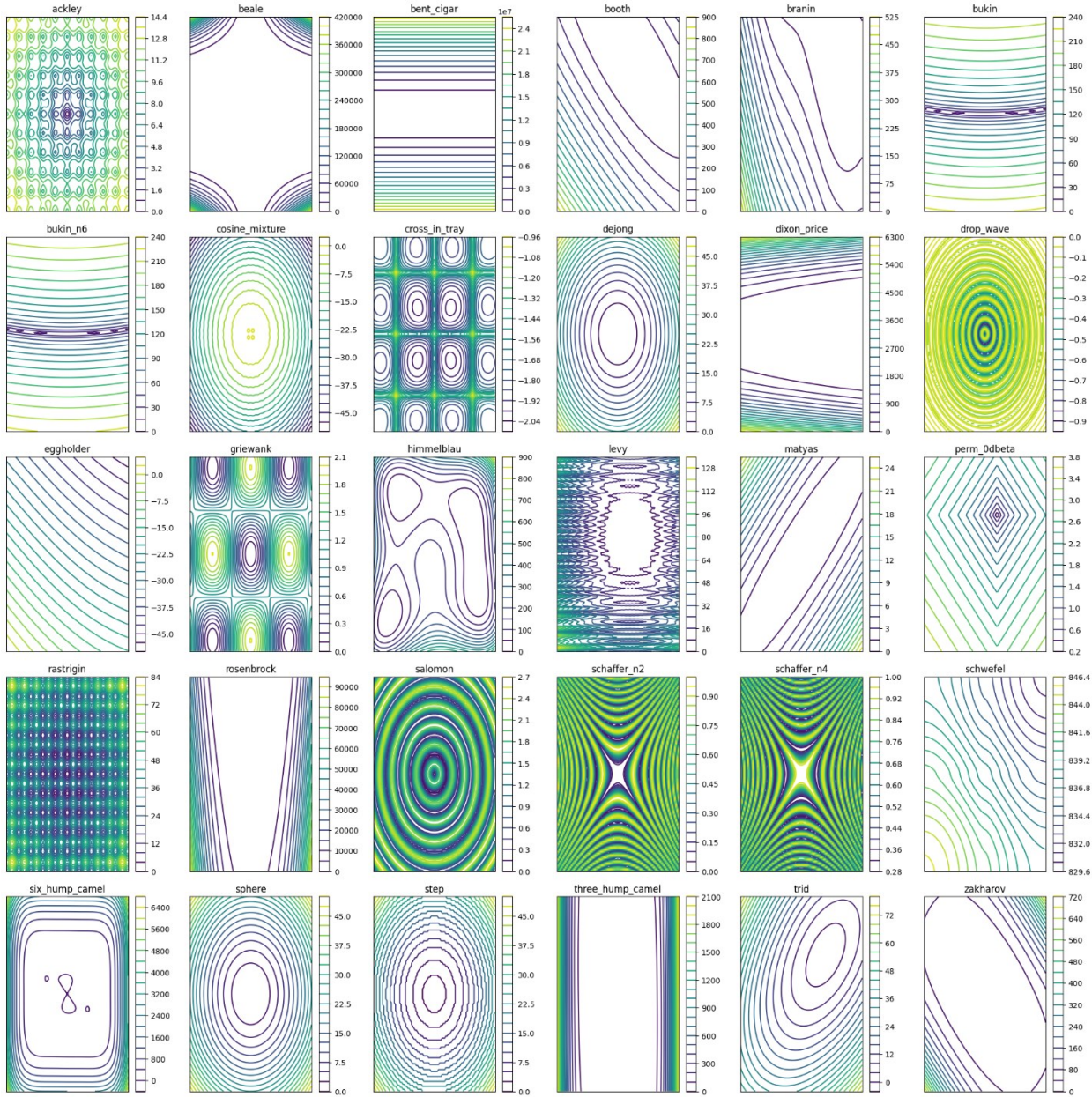

Fig. 1 Visualization of the single objective benchmark functions

The outcome of the benchmarking analysis on the single objective functions above is listed in Table 4. Overall, SPINEX can completely minimize the examined functions and rank first (except for Bukin and Bukin N.6, where it ranks third). In terms of execution time, SPINEX has varying ranks, achieving first in a few functions. Comparatively, and when compared among all functions, SPINEX ranks first in fitness and sixth in execution time. Table 5 further lists the ranking of the other algorithms.

Table 5 Overall ranking on the mathematical benchmarking functions

| Algorithm | Total Fitness Rank Sum | Overall Ranking | Time Rank Sum | Time Ranking |
|---|---|---|---|---|
| SPINEX | 35 | 1 | 181 | 6 |
| Differential Evolution | 38 | 2 | 311 | 11 |
| Basin-hopping | 38 | 3 | 152 | 4 |
| Nelder-Mead | 49 | 4 | 41 | 1 |
| Bat Algorithm | 87 | 5 | 206 | 7 |
| Flower Pollination Algorithm | 131 | 6 | 242 | 8 |
| Cuckoo Search | 196 | 7 | 165 | 5 |
| Grey Wolf Optimizer | 219 | 8 | 285 | 9 |
| Biogeography-Based Optimization | 252 | 9 | 303 | 10 |
| Harmony Search | 290 | 10 | 89 | 3 |
| SA | 300 | 11 | 62 | 2 |

Table 4 Comparison of results with regard to the mathematical benchmarking functions

| Function | SPINEX Fitness | Rank | Time | Rank | SA Fitness | Rank | Time | Rank | DE Fitness | Rank | Time | Rank | BH Fitness | Rank | Time | Rank | NMA Fitness | Rank | Time | Rank | CS Fitness | Rank | Time | Rank |
|---|---|---|---|---|---|---|---|---|---|---|---|---|---|---|---|---|---|---|---|---|---|---|---|---|
| Ackley | 0 | 1 | 0.105746 | 4 | 1.446601 | 11 | 0.005955 | 2 | 0 | 1 | 32.813684 | 11 | 0 | 1 | 1.247291 | 6 | 0 | 1 | 0.001024 | 1 | 0.007344 | 8 | 0.577499 | 5 |
| Beale | 0 | 1 | 0.004987 | 4 | 0.016933 | 10 | 0.003012 | 1 | 0 | 1 | 26.669554 | 11 | 0 | 1 | 0.310414 | 6 | 0 | 1 | 0.003989 | 3 | 0.000007 | 5 | 0.136815 | 5 |
| Bent Cigar | 0 | 1 | 0.00399 | 4 | 81.282227 | 11 | 0.003325 | 3 | 0 | 1 | 25.066676 | 11 | 0 | 1 | 0.308866 | 6 | 0 | 1 | 0.000996 | 1 | 1.826109 | 6 | 0.109706 | 5 |
| Booth | 0 | 1 | 0.002992 | 1 | 0.008117 | 9 | 0.002992 | 1 | 0 | 1 | 27.016599 | 11 | 0 | 1 | 0.081781 | 5 | 0 | 1 | 0.002992 | 1 | 0.000019 | 7 | 0.121849 | 6 |
| Booth | 0 | 1 | 0.002992 | 1 | 0.008117 | 9 | 0.002992 | 1 | 0 | 1 | 27.016599 | 11 | 0 | 1 | 0.081781 | 5 | 0 | 1 | 0.002992 | 1 | 0.000019 | 7 | 0.121849 | 6 |
| Branin | 0.397887 | 1 | 5.300663 | 10 | 0.532609 | 11 | 0.004016 | 3 | 0.397887 | 1 | 3.945012 | 7 | 0.397887 | 1 | 0.123885 | 4 | 0.397887 | 1 | 0.003989 | 3 | 0.397909 | 8 | 0.162819 | 5 |
| Bukin | 0.1 | 3 | 6.539021 | 10 | 8.319309 | 11 | 0.002992 | 2 | 0.092914 | 2 | 25.864903 | 11 | 0.070452 | 1 | 0.992507 | 5 | 0.1 | 3 | 0.002029 | 1 | 1.109794 | 9 | 0.182307 | 4 |
| Bukin N.6 | 0.1 | 3 | 5.413713 | 9 | 13.956005 | 11 | 0.003992 | 2 | 0.073167 | 1 | 26.706208 | 11 | 0.088699 | 2 | 0.870896 | 5 | 0.1 | 3 | 0.001994 | 1 | 1.355188 | 8 | 0.174465 | 4 |
| Cosine Mixture | -7.08719E+41 | 1 | 6.596196 | 10 | -52.490603 | 3 | 0.003989 | 2 | -52.490603 | 3 | 8.745412 | 11 | -52.490603 | 3 | 0.04391 | 4 | -0.2 | 11 | 0.001001 | 1 | -52.490603 | 3 | 0.16761 | 5 |
| Cross-in-Tray | -2.062612 | 1 | 5.625896 | 9 | -2.057572 | 11 | 0.00299 | 2 | -2.062612 | 1 | 3.369174 | 7 | -2.062612 | 1 | 0.093601 | 4 | -2.062612 | 1 | 0.003989 | 2 | -2.06261 | 9 | 0.235351 | 5 |
| DeJong | 0 | 1 | 0.004986 | 4 | 0.074889 | 11 | 0.002992 | 2 | 0 | 1 | 27.688305 | 11 | 0 | 1 | 0.069844 | 5 | 0 | 1 | 0 | 1 | 0.000006 | 7 | 0.149752 | 6 |
| Dixon-Price | 0 | 1 | 0.002992 | 3 | 0.019902 | 10 | 0.002993 | 4 | 0 | 1 | 27.67926 | 11 | 0 | 1 | 0.1652 | 6 | 0 | 1 | 0.002021 | 1 | 0.000038 | 7 | 0.122977 | 5 |
| Drop-Wave | -1 | 1 | 5.445123 | 9 | -0.143442 | 11 | 0.003021 | 2 | -1 | 1 | 9.081832 | 11 | -1 | 1 | 0.125694 | 4 | -1 | 1 | 0 | 1 | -0.936245 | 8 | 0.177784 | 5 |
| Eggholder | -66.843717 | 1 | 6.353093 | 11 | -49.489936 | 3 | 0.002989 | 2 | -49.489936 | 3 | 5.074628 | 8 | -49.489936 | 3 | 0.047379 | 4 | -49.489936 | 3 | 0.000998 | 1 | -49.489936 | 3 | 0.241694 | 5 |
| Griewank | 0 | 1 | 0.01496 | 4 | 0.090606 | 11 | 0.00496 | 3 | 0 | 1 | 20.105166 | 11 | 0 | 1 | 0.102727 | 5 | 0 | 1 | 0 | 1 | 0.000001 | 5 | 0.419257 | 6 |
| Himmelblau | 0 | 1 | 0.00302 | 3 | 0.010389 | 10 | 0.002991 | 2 | 0 | 1 | 27.196104 | 11 | 0 | 1 | 0.127341 | 6 | 0 | 1 | 0.005015 | 4 | 0.000047 | 6 | 0.121704 | 5 |
| Levy | 0 | 1 | 0.022913 | 3 | 0.219369 | 11 | 0.005013 | 1 | 0 | 1 | 30.917695 | 11 | 0 | 1 | 0.229413 | 5 | 0 | 1 | 0.006331 | 2 | 0.000007 | 8 | 0.650097 | 6 |
| Matyas | 0 | 1 | 0.002992 | 2 | 0.069492 | 11 | 0.002996 | 4 | 0 | 1 | 25.708067 | 11 | 0 | 1 | 0.094534 | 5 | 0 | 1 | 0.000999 | 1 | 0.000001 | 7 | 0.117711 | 6 |
| Perm 0,d,beta | -7 | 1 | 6.981791 | 11 | -6.836098 | 11 | 0.002993 | 1 | -7 | 1 | 3.32124 | 7 | -7 | 1 | 0.110185 | 4 | -7 | 1 | 0.003989 | 3 | -6.999999 | 7 | 0.175558 | 5 |
| Rastrigin | 0 | 1 | 0.010999 | 4 | 2.710004 | 11 | 0.003961 | 2 | 0 | 1 | 19.78841 | 11 | 0 | 1 | 0.19052 | 5 | 0 | 1 | 0.000996 | 1 | 0.000402 | 6 | 0.352171 | 6 |
| Rosenbrock | 0 | 1 | 0.013963 | 4 | 0.574613 | 11 | 0.006009 | 2 | 0 | 1 | 29.197395 | 11 | 0 | 1 | 0.346997 | 5 | 0 | 1 | 0.00599 | 1 | 0.000402 | 7 | 0.372638 | 6 |
| Salomon | 0 | 1 | 0.062375 | 4 | 0.195326 | 11 | 0.002991 | 2 | 0 | 1 | 27.451686 | 11 | 0 | 1 | 0.150595 | 5 | 0 | 1 | 0 | 1 | 0.000046 | 5 | 0.183538 | 6 |
| Schaffer N.2 | 0 | 1 | 0.00402 | 3 | 0.082332 | 11 | 0.003961 | 2 | 0 | 1 | 21.208246 | 11 | 0 | 1 | 0.226786 | 6 | 0 | 1 | 0.000971 | 1 | 0 | 1 | 0.158396 | 5 |
| Schaffer N.4 | 0.292579 | 1 | 6.386757 | 10 | 0.324918 | 11 | 0.002991 | 1 | 0.292579 | 1 | 13.375 | 11 | 0.292579 | 1 | 0.292536 | 5 | 0.292579 | 1 | 0.004987 | 2 | 0.292587 | 9 | 0.195006 | 4 |
| Schwefel | 830.075197 | 1 | 5.739069 | 9 | 830.080936 | 3 | 0.003989 | 2 | 830.080936 | 3 | 0.29575 | 6 | 830.080936 | 3 | 0.056821 | 4 | 830.080936 | 3 | 0.002023 | 1 | 830.080936 | 3 | 0.263958 | 5 |
| Six-Hump Camel | -1.031628 | 1 | 5.728368 | 11 | -0.801942 | 11 | 0.004987 | 3 | -1.031628 | 1 | 3.531689 | 7 | -1.031628 | 1 | 0.133703 | 4 | -1.031628 | 1 | 0.003018 | 1 | -1.031589 | 7 | 0.137327 | 5 |
| Sphere | 0 | 1 | 0.006947 | 3 | 0.097269 | 11 | 0.003989 | 2 | 0 | 1 | 26.879531 | 11 | 0 | 1 | 0.069843 | 5 | 0 | 1 | 0.000998 | 1 | 0.000013 | 7 | 0.225397 | 6 |
| Step | 0 | 1 | 0.005019 | 4 | 0 | 1 | 0.003985 | 2 | 0 | 1 | 4.342366 | 9 | 0 | 1 | 0.03796 | 5 | 0 | 1 | 0 | 1 | 0 | 1 | 0.178706 | 6 |
| Three-Hump Camel | 0 | 1 | 0.002992 | 2 | 0.305341 | 11 | 0.002992 | 2 | 0 | 1 | 28.214526 | 11 | 0 | 1 | 0.106879 | 5 | 0 | 1 | 0 | 1 | 0.000002 | 7 | 0.13846 | 6 |
| Trid | -2 | 1 | 6.170678 | 11 | -1.797021 | 11 | 0.002992 | 2 | -2 | 1 | 2.931686 | 7 | -2 | 1 | 0.088381 | 4 | -2 | 1 | 0.001991 | 1 | -1.999993 | 7 | 0.161563 | 5 |
| Zakharov | 0 | 1 | 0.013963 | 4 | 0.428166 | 11 | 0.005013 | 2 | 0 | 1 | 28.977818 | 11 | 0 | 1 | 0.128655 | 5 | 0 | 1 | 0 | 1 | 0.000001 | 7 | 0.409471 | 6 |

| Function | GWO Fitness | Rank | Time | Rank | BA Fitness | Rank | Time | Rank | HS Fitness | Rank | Time | Rank | FPA Fitness | Rank | Time | Rank | BBO Fitness | Rank | Time | Rank |
|---|---|---|---|---|---|---|---|---|---|---|---|---|---|---|---|---|---|---|---|---|
| Ackley | 0.00312 | 7 | 10.151963 | 10 | 0.000083 | 5 | 3.822096 | 7 | 0.581568 | 10 | 0.019975 | 3 | 0.000613 | 6 | 6.111429 | 8 | 0.297472 | 9 | 7.643759 | 9 |
| Beale | 0.000015 | 7 | 4.417917 | 9 | 0 | 1 | 1.694298 | 7 | 0.09851 | 11 | 0.003031 | 2 | 0.000015 | 7 | 3.899611 | 8 | 0.001109 | 9 | 5.357256 | 10 |
| Bent Cigar | 2.582981 | 7 | 4.165976 | 9 | 17.276442 | 9 | 1.672691 | 7 | 56.626849 | 10 | 0.001995 | 2 | 0.000012 | 5 | 3.942564 | 8 | 3.841075 | 8 | 5.462857 | 10 |
| Booth | 0.104835 | 11 | 4.262391 | 9 | 0 | 1 | 1.619835 | 7 | 0.044606 | 10 | 0.00302 | 4 | 0.000002 | 6 | 3.925892 | 8 | 0.001857 | 8 | 5.50923 | 10 |
| Booth | 0.104835 | 11 | 4.262391 | 9 | 0 | 1 | 1.619835 | 7 | 0.044606 | 10 | 0.00302 | 4 | 0.000002 | 6 | 3.925892 | 8 | 0.001857 | 8 | 5.50923 | 10 |
| Branin | 0.397892 | 7 | 4.81618 | 9 | 0.397887 | 1 | 1.838449 | 6 | 0.457006 | 10 | 0.003989 | 1 | 0.397887 | 1 | 4.156445 | 8 | 0.403114 | 9 | 5.501069 | 11 |
| Bukin | 0.124429 | 5 | 5.11225 | 8 | 0.143458 | 6 | 1.93787 | 6 | 0.387154 | 8 | 0.005016 | 3 | 0.232605 | 7 | 4.338092 | 7 | 3.638621 | 10 | 6.015066 | 9 |
| Bukin N.6 | 0.117605 | 5 | 4.843755 | 8 | 0.154095 | 6 | 1.814995 | 6 | 4.466948 | 10 | 0.004986 | 3 | 0.161496 | 7 | 4.053109 | 7 | 1.536521 | 9 | 5.488996 | 10 |
| Cosine Mixture | -52.490603 | 3 | 4.845597 | 8 | -74.8437 | 2 | 1.915184 | 6 | -50.133913 | 10 | 0.00399 | 3 | -52.490603 | 3 | 4.074877 | 7 | -52.204975 | 9 | 5.583144 | 9 |
| Cross-in-Tray | -2.062611 | 7 | 5.712571 | 10 | -2.062612 | 1 | 2.119205 | 6 | -2.061331 | 10 | 0.006981 | 3 | -2.062612 | 1 | 4.391864 | 8 | -2.062611 | 7 | 5.992649 | 11 |
| DeJong | 0.003992 | 8 | 4.795491 | 9 | 0 | 1 | 1.841146 | 7 | 0.00564 | 9 | 0.004017 | 3 | 0 | 1 | 4.127009 | 8 | 0.00726 | 10 | 5.582822 | 10 |
| Dixon-Price | 0.000352 | 8 | 4.304848 | 9 | 0 | 1 | 1.684647 | 7 | 0.093592 | 11 | 0.002983 | 2 | 0.000007 | 6 | 3.891091 | 8 | 0.001569 | 9 | 5.371022 | 10 |
| Drop-Wave | -0.999972 | 5 | 4.898881 | 8 | -0.936245 | 8 | 1.821645 | 6 | -0.932111 | 10 | 0.004763 | 3 | -0.997131 | 6 | 4.147765 | 7 | -0.973994 | 7 | 5.525429 | 10 |
| Eggholder | -49.265717 | 10 | 5.729978 | 9 | -52.877514 | 2 | 2.152192 | 6 | -48.701194 | 11 | 0.00698 | 3 | -49.489936 | 3 | 4.409696 | 7 | -49.424539 | 9 | 5.984929 | 10 |
| Griewank | 0.000524 | 8 | 8.232424 | 10 | 0.007396 | 10 | 3.001642 | 7 | 0.000242 | 7 | 0.013963 | 3 | 0.000026 | 6 | 5.121969 | 8 | 0.000764 | 9 | 6.862468 | 9 |
| Himmelblau | 0.007102 | 9 | 4.277651 | 9 | 0 | 1 | 1.643492 | 7 | 0.29263 | 11 | 0.002964 | 1 | 0.000049 | 7 | 3.850461 | 8 | 0.002517 | 8 | 5.345909 | 10 |
| Levy | 0 | 1 | 10.681896 | 10 | 0 | 1 | 3.955394 | 7 | 0.005499 | 10 | 0.022967 | 4 | 0 | 1 | 6.280588 | 8 | 0.000426 | 9 | 7.787991 | 9 |
| Matyas | 0 | 1 | 4.075894 | 9 | 0 | 1 | 1.592391 | 7 | 0.000564 | 10 | 0.002992 | 2 | 0.000001 | 7 | 3.779085 | 8 | 0.000219 | 9 | 5.223431 | 10 |
| Perm 0,d,beta | -6.994515 | 9 | 5.236559 | 9 | -7 | 1 | 1.990133 | 6 | -6.963942 | 10 | 0.00385 | 2 | -7 | 1 | 4.377401 | 8 | -6.996846 | 8 | 5.801109 | 10 |
| Rastrigin | 0.003287 | 8 | 6.855516 | 10 | 0 | 1 | 2.572526 | 7 | 1.241541 | 10 | 0.01 | 3 | 0.001294 | 7 | 4.92678 | 8 | 0.373895 | 9 | 6.611624 | 9 |
| Rosenbrock | 0.000118 | 5 | 7.393202 | 10 | 0.028814 | 9 | 2.828891 | 7 | 0.000238 | 6 | 0.012967 | 3 | 0.000494 | 8 | 4.934548 | 8 | 0.0813 | 10 | 6.735165 | 9 |
| Salomon | 0.083753 | 8 | 5.017317 | 9 | 0.099873 | 9 | 1.873336 | 7 | 0.102445 | 10 | 0.005989 | 3 | 0.000563 | 6 | 4.135152 | 8 | 0.017295 | 7 | 5.592525 | 10 |
| Schaffer N.2 | 0.000152 | 10 | 4.807052 | 9 | 0 | 1 | 1.819092 | 7 | 0.000041 | 9 | 0.004551 | 4 | 0 | 1 | 4.14315 | 8 | 0.000004 | 8 | 5.693732 | 10 |
| Schaffer N.4 | 0.292579 | 1 | 5.205773 | 8 | 0.292579 | 1 | 1.963933 | 6 | 0.292722 | 10 | 0.005321 | 3 | 0.292585 | 8 | 4.271755 | 7 | 0.292579 | 1 | 5.71888 | 9 |
| Schwefel | 830.202872 | 11 | 6.270799 | 11 | 830.075197 | 1 | 2.339213 | 7 | 830.138846 | 10 | 0.007951 | 3 | 830.080936 | 3 | 4.513196 | 8 | 830.087332 | 9 | 6.13818 | 10 |
| Six-Hump Camel | -0.885238 | 10 | 4.473955 | 9 | -1.031628 | 1 | 1.735329 | 6 | -1.016819 | 9 | 0.004016 | 2 | -1.031627 | 6 | 3.972949 | 8 | -1.023251 | 8 | 5.66082 | 10 |
| Sphere | 0.000195 | 9 | 6.07376 | 10 | 0 | 1 | 2.213272 | 7 | 0.000083 | 8 | 0.006954 | 4 | 0 | 1 | 4.334723 | 8 | 0.001296 | 10 | 5.88976 | 9 |
| Step | 0 | 1 | 5.241434 | 10 | 0 | 1 | 1.977909 | 7 | 0 | 1 | 0.004994 | 3 | 0 | 1 | 4.288731 | 8 | 0 | 1 | 5.781414 | 11 |
| Three-Hump Camel | 0.061043 | 10 | 4.384464 | 9 | 0 | 1 | 1.74994 | 7 | 0.023874 | 9 | 0.003251 | 4 | 0 | 1 | 4.099369 | 8 | 0.001047 | 8 | 5.533215 | 10 |
| Trid | -1.995631 | 9 | 4.952718 | 9 | -2 | 1 | 1.88179 | 6 | -1.946572 | 10 | 0.00399 | 3 | -2 | 1 | 4.245568 | 8 | -1.999824 | 8 | 5.720295 | 10 |
| Zakharov | 0.000006 | 8 | 8.004899 | 10 | 0 | 1 | 2.952788 | 7 | 0.050503 | 10 | 0.012965 | 3 | 0 | 1 | 5.031329 | 8 | 0.001868 | 9 | 6.793357 | 9 |

Then, fifteen functions were extended into multiple and many objectives. These functions include Ackley, Alpine, Bohachevsky, Dixon-Price, Griewank, Levy, Michalewicz, Qing, Rosenbrock, Salomon, Schaffer, Sphere, Styblinski-Tang, and Zakharov. These functions were examined at objective counts of 2, 5, 10, 25, and 50, dimensions of 2, 5, 10, 25, and 50, and population sizes of 100 and 1000. Given the large number of objectives, dimensions, and population sizes evaluated in this leg of the experiments, the detailed tables for each case will be provided as supplementary material. For each test function, dimension, and population size, algorithms are ranked based on their performance in each objective and execution time. These rankings are summed to produce total objective and time rank sums, which determine overall and time rankings. This system is then extended to calculate average rankings across all experiments, resulting in final overall and time rankings. This process is repeated for all experimental configurations, and the results are aggregated to calculate average rankings. Table 6 shows that SPINEX ranks second and third in terms of its performance under fitness and execution time.

Table 6 Ranking across experiments

| Algorithm | Avg Overall Ranking | Final Overall Ranking | Avg Time Ranking | Final Time Ranking |
|---|---|---|---|---|
| UNSGA3 | 3.38 | 1 | 4.90 | 4 |
| SPINEX | 3.84 | 2 | 4.46 | 3 |
| NSGA-III | 4.20 | 3 | 5.76 | 8 |
| NSGA-II | 4.58 | 4 | 5.72 | 7 |
| R-NSGA-II | 4.92 | 5 | 5.68 | 6 |
| AGE-MOEA | 5.140 | 6 | 3.48 | 2 |
| SMS-EMOA | 5.220 | 7 | 7.66 | 9 |
| SPEA2 | 6.20 | 8 | 5.40 | 5 |
| MOEA/D | 7.520 | 9 | 1.64 | 1 |

To allow for a more comprehensive evaluation of the examined algorithms, Fig. 2 shows a comparative performance of the average fitness across different dimensions and execution time across dimensions, objectives, and population sizes. In this plot, the horizontal axis represents the number of dimensions, while the vertical shows the mean fitness value across all objectives. For each algorithm, the plot displays a line that tracks how its average fitness changes with increasing dimensions. The fitness values are calculated by taking the mean of all objective values for each algorithm at each dimension (and hence, may slightly differ from the tabular rankings). Lower fitness values typically indicate better performance in optimization problems. As one can see, SPINEX is stable regarding fitness and execution time across dimensions and population sizes. The execution time scales comparatively to other algorithms in the case of execution time across various objective numbers.

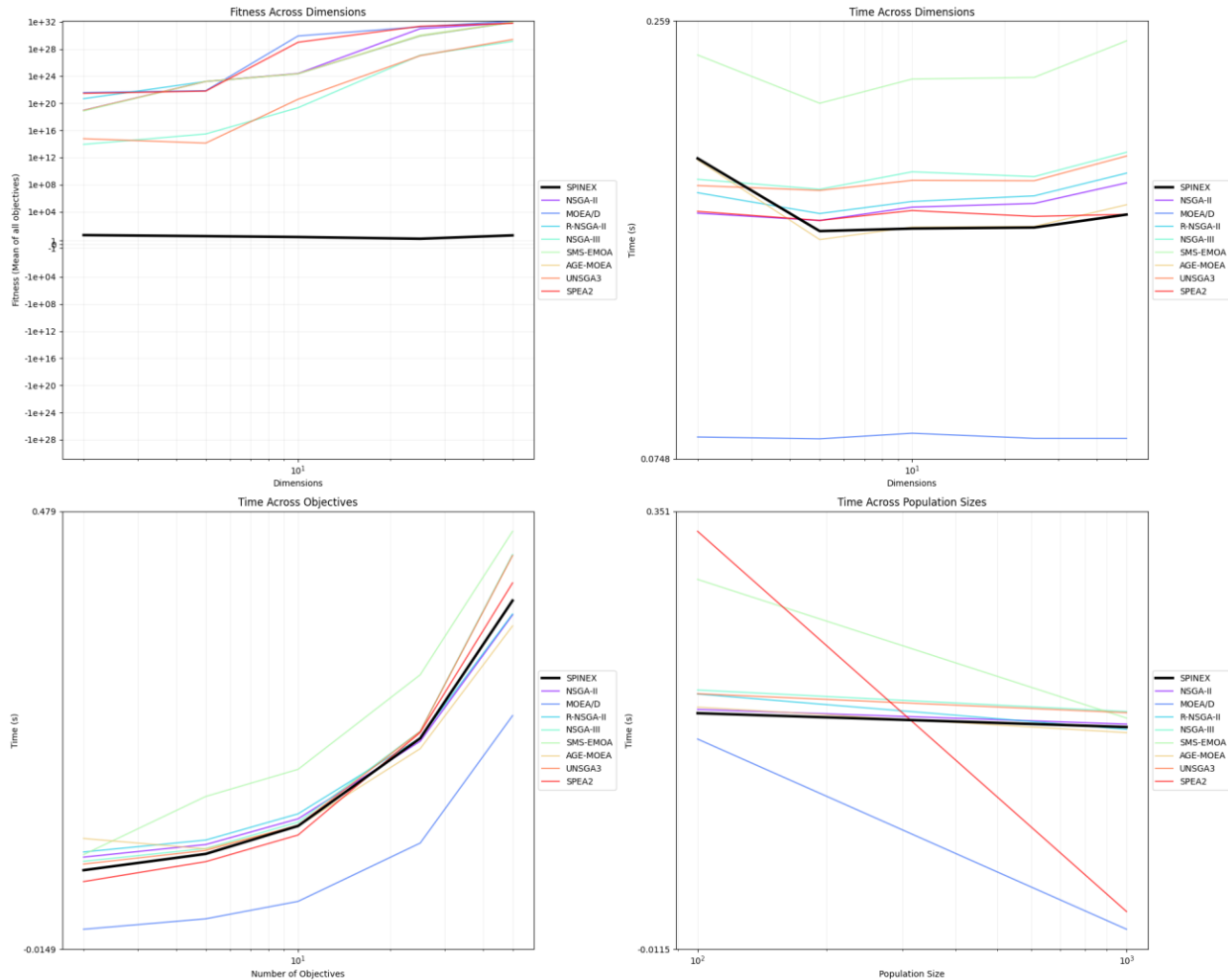

Fig. 2 Further findings

*4.2 Realistic optimization scenarios*

Five new realistic optimization scenarios were created to further examine the performance of SPINEX. While Table 7 lists a summary of these scenarios, the following presents additional details.

Table 7 Realistic scenarios

| Function | Single Objective | Multi- & Many-Objective |
|---|---|---|
| ML Hyperparameter Tuning | Minimizes training time to accuracy ratio | Optimizes accuracy, complexity, training time, memory usage, inference time, and generalization [6 objectives] |
| Supply Chain Management | Minimizes cost-to-service level ratio | Optimizes inventory cost, ordering cost, transportation cost, quality cost, carbon footprint, service level, and resilience [7 objectives] |
| Sustainable City Planning | Minimizes cost-to-sustainability score ratio | Optimizes energy consumption, green space, traffic congestion, air quality, water consumption, public transport usage, waste generation, employment rate, cost of living, and citizen satisfaction [10 objectives] |
| Traffic Signal Timing | Minimizes wait time to throughput ratio | Optimizes average wait time, maximum wait time, and throughput [3 objectives] |
| Truss Design | Minimizes cost-to-strength ratio | Optimizes strength-to-weight ratio, cost, and stability [3 objectives] |

### 4.2.1 Truss design optimization

This classic engineering challenge aims to create the most efficient structural truss support system. Both the single-objective and multi-objective versions of this function focus on optimizing the design of a truss, but they differ in their approach and complexity. The single objective version focuses on minimizing the cost-to-strength ratio by taking into account two primary variables: length and thickness. The function scales these inputs to realistic ranges (0-5 meters for length and 0-0.1 meters for thickness). The strength is calculated as a simple product of length and thickness multiplied by a constant factor. The weight is determined using the density of steel (7850 kg/m³), and the cost includes both material cost and a fixed manufacturing cost. The objective is to minimize the ratio of cost to strength, effectively seeking the most cost-effective design. The objective function is represented as:

$$Objective = \frac{Cost}{Strength}$$

where,
- $Cost = 5 \times Weight + 1000 \ (material\ cost\ plus\ a\ fixed\ manufacturing\ cost)$
- $Weight = 7850 \times Length \times Thickness \ (based\ on\ the\ density\ of\ steel)$
- $Strength = 1000 \times Length \times Thickness$

In contrast, the multi-objective version considers additional factors by including support beams, which adds more variables to the optimization problem. This version optimizes three objectives simultaneously:
1. Maximizing the strength-to-weight ratio
2. Minimizing the total cost
3. Maximizing the smallest dimension for stability

$$Objective\ 1 = \frac{Strength}{Weight}$$
$$Objective\ 2 = Total\ cost$$
$$Objective\ 3 = -Minimum\ Dimension$$

Where:
- $Strength\ = 1000 \times Main\ Length \times Main\ Thickness + Support\ Strength$
- $Weight\ = 7850 \times Main\ Length \times Main\ Thickness + Support\ Weight$
- $Total\ Cost = Material\ Cost + Manufacturing\ Cost$

This function scales the inputs differently, allowing for finer control over the main beam dimensions (0-5 meters for length and 0-0.5 meters for thickness). It also incorporates support beams when more than two variables are provided, scaling them to 0-3 meters. The strength calculation considers both the main beam and support beams separately.

### 4.2.2 ML hyperparameter tuning

Machine learning hyperparameter tuning for developing effective ML models. This function's single objective and multi-objective versions represent two approaches to this optimization problem. For example, the single objective version focuses on optimizing the ratio of training time to simulated accuracy by considering three hyperparameters:
1. Learning rate (scaled from $1e^{-5}$ to $1e^{-3}$)
2. Number of layers (1 to 10)
3. Neurons per layer (10 to 100)

$$Objective = \frac{Training\ Time}{Simulated\ accuracy}$$

where,
- $Training\ Time = Complexity \times \log(\frac{1}{Learning\ rate})$
- $Simulated\ Accuracy = 1 - \exp(\frac{-Learning\ rate\ \times\ Complexity}{1000})$
- $Complexity = Num\ Layers \times Neurons\ Per\ Layer$

The function calculates a complexity measure based on the number of layers and neurons to estimate the model's accuracy and training time. The objective is to minimize the training time to accuracy ratio and seek the most efficient model regarding training speed and performance. The multi-objective version is more complex and aims to optimize the following six objectives simultaneously:

$Objective\ 1 = -Simulated\ Accuracy$

$Objective\ 2 = Model\ Complexity$

$Objective\ 3 = Training\ Time$

$Objective\ 4 = Memory\ Usage$

$Objective\ 5 = Inference\ Time$

$Objective\ 6 = Generalization\ Gap$

### 4.2.3 Supply chain management

Supply chain management is a complex field that involves balancing multiple competing objectives to ensure efficient and effective distribution of goods. In this function, the objective variant focuses on optimizing the total cost-to-service level ratio. It considers the following factors:

1. Order quantities for each product (0 to 1000 units)
2. Holding cost (proportional to order quantities)
3. Fixed ordering cost per product
4. Transportation cost (fixed component plus variable based on order quantities)
5. Service level (calculated based on mean order quantity)

$$Objective = \frac{Total\ Cost}{Service\ level}$$

where,
- $Total\ Cost = Holding\ Cost + Ordering\ Cost + Transportation\ Cost$
- $Service\ level = 1 - \exp(\frac{-Mean\ Order\ Quantities}{500})$

The objective is to minimize the ratio of total cost to service level to seek the most cost-effective supply chain configuration that maintains a high level of service. Similarly, the multi-objective version optimizes seven objectives simultaneously:

*Objective 1=Inventory Cost*

*Objective 2=Ordering Cost*

*Objective 3=Transportation Cost*

*Objective 4=Quality Cost*

*Objective 5=Carbon Footprint*

*Objective 6=Service Level Score*

*Objective 7=Resilience Score*

### 4.2.4 Traffic signal timing

Traffic signal timing is a critical aspect of urban planning and traffic management that often optimizes the flow of vehicles and pedestrians through intersections. From this lens, the single-objective version focuses on optimizing the ratio of average wait time to throughput by considering the following factors:
1. Green times for each phase of the traffic signal (0 to 60 seconds)
2. Cycle time (sum of all green times)
3. Average wait time (assumed to be half the cycle time)
4. Throughput (assumed to be inversely proportional to cycle time)

$$Objective = \frac{Average\ Wait\ Time}{Throughput}$$

where,
- $Average\ Wait\ Time = \frac{Cycle\ time}{2}$
- $Throughput = \frac{1000}{Cycle\ time}$

The multi-objective version considers three separate objectives. By separating average and maximum wait times, this variant can identify solutions that improve overall efficiency and prevent excessive delays for any particular direction of traffic.

$Objective\ 1 = Average\ Wait\ Time$

$Objective\ 2 = Maximum\ Wait\ Time$

$Objective\ 3 = -Throughput$

### 4.2.5 Sustainable city planning

Sustainable city planning is a multifaceted challenge that aims to create urban environments that are environmentally friendly, economically viable, and socially beneficial. This function's single objective and multi-objective versions represent two different approaches to this optimization problem. The single objective version focuses on optimizing the ratio of implementation cost to sustainability score. The sustainability score is calculated as the average of these three factors, while the cost is assumed to be directly proportional to the sum of the input values. The objective is to minimize the cost to sustainability score ratio, effectively seeking the most cost-effective way to improve the city's sustainability. This score has three main objectives:
1. Energy efficiency (0-100%)
2. Green space (0-30% of city area)
3. Public transport usage (0-50%)

$$Objective = \frac{Cost}{Sustainability\ score}$$

where,

- Sustainability score = $\frac{Energy efficiency + Green\ Space + Public\ Transport}{2}$
- $Cost = 1000 \times Energy efficiency + Green\ Space + Public\ Transport)$

The multi-objective version is significantly more complex and considers ten objectives simultaneously:

$Objective\ 1 = Energy\ Consumption$
$Objective\ 2 = -Green\ Space$
$Objective\ 3 = Traffic\ Congestion$
$Objective\ 4 = Air\ Quality$
$Objective\ 5 = Water\ Consumption$
$Objective\ 6 = -Public\ Transport$
$Objective\ 7 = Waste\ Generation$
$Objective\ 8 = -Employment\ Rate$
$Objective\ 9 = Cost\ of\ Living$
$Objective\ 10 = -Citizen\ Satisfaction$

Table 8 lists the fitness and execution time rankings for all algorithms as gathered from the aforementioned optimization functions. It is quite clear that SPINEX ranks first and third under fitness and eighth and third under execution time for the single and multi-optimization scenarios, respectively.

Table 8 Overall ranking

| Algorithm | Single Objective Overall Ranking | Single Objective Time Ranking | Algorithm | Multi Objectives Overall Ranking | Multi Objectives Time Ranking |
|---|---|---|---|---|---|
| SPINEX | 1 | 8 | AGE-MOEA | 1 | 6 |
| BH | 2 | 5 | SMS-EMOA | 2 | 9 |
| FPA | 3 | 9 | SPINEX | 3 | 8 |
| CS | 4 | 4 | R-NSGA-II | 4 | 5 |
| DE | 5 | 7 | SPEA2 | 5 | 7 |
| NMA | 6 | 1 | UNSGA3 | 6 | 1 |
| BA | 7 | 6 | MOEA/D | 7 | 3 |
| GWO | 8 | 10 | NSGA-III | 7 | 2 |
| HS | 9 | 3 | NSGA-II | 9 | 4 |
| SA | 10 | 2 | | | |
| BBO | 11 | 11 | | | |

As one can see, SPINEX presents a competitive performance in both types of scenarios. For example, this algorithm continues to achieve high fitness performance across the different dimensions and population sizes. Its performance in terms of exaction time is good and seems to lie in the middle of the other algorithms. To showcase a deeper granularity analysis with respect to the truss design scenario and sustainable city planning scenario, Fig. 3 traces the behavior of SPINEX and other algorithms. This figure shows a consistent behavior from SPINEX, wherein it tends to perform well in fitness calculation and is slightly costly on the exaction time front.

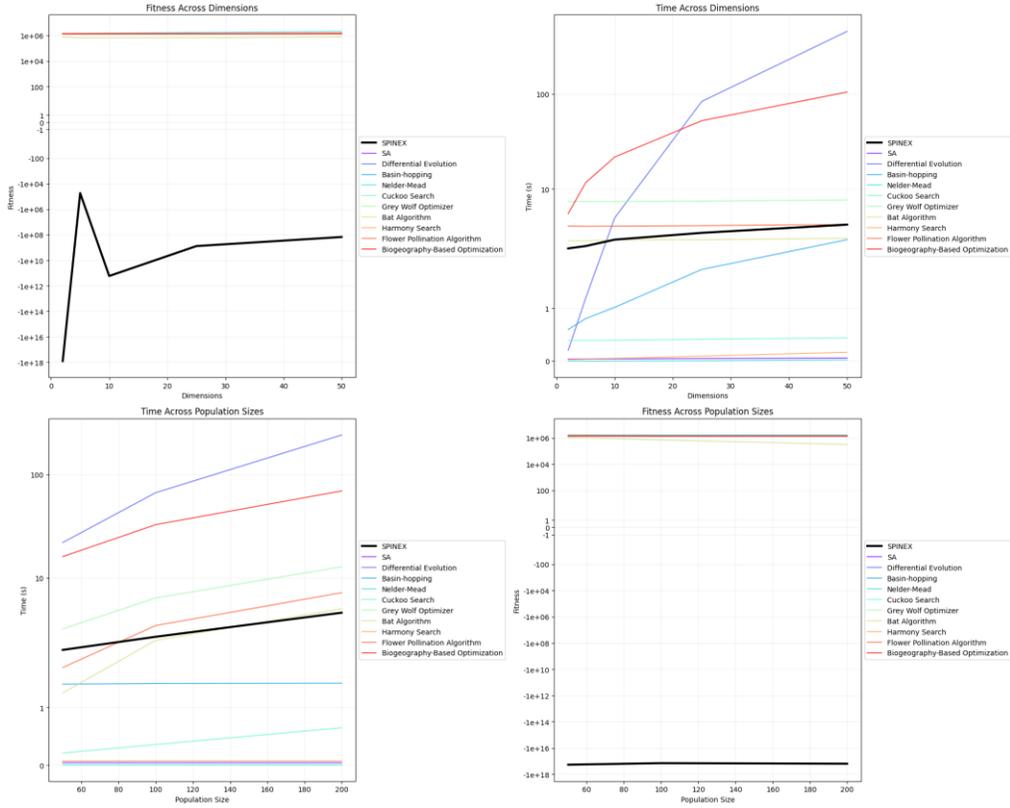

(a) Single objective

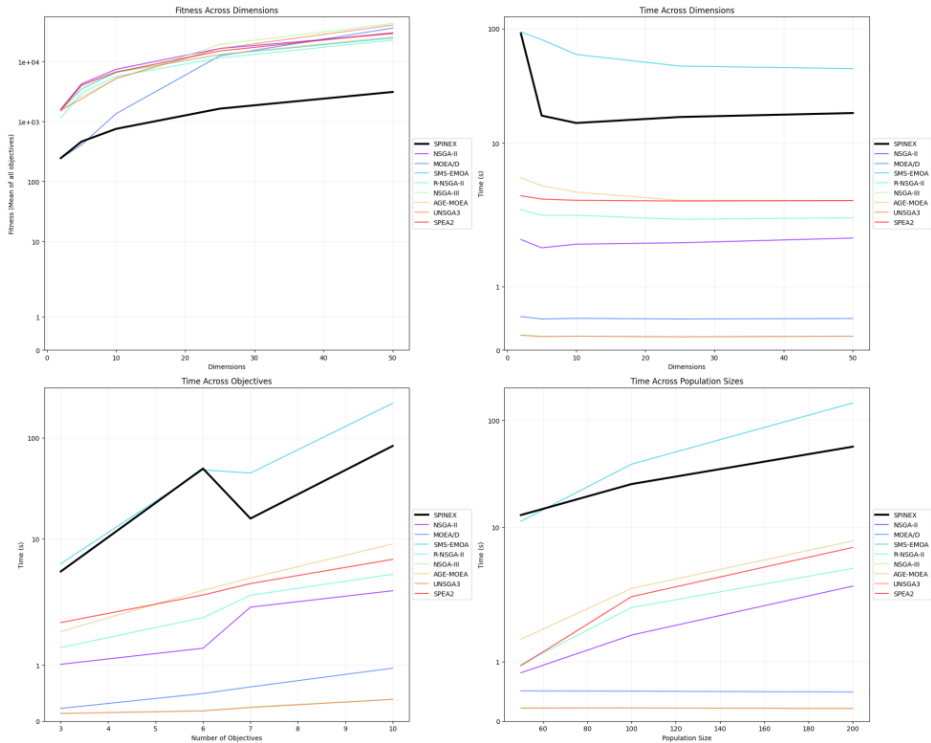

(b) Multi and multiple objectives

Fig. 3 Additional findings across dimensions, objectives, and population sizes

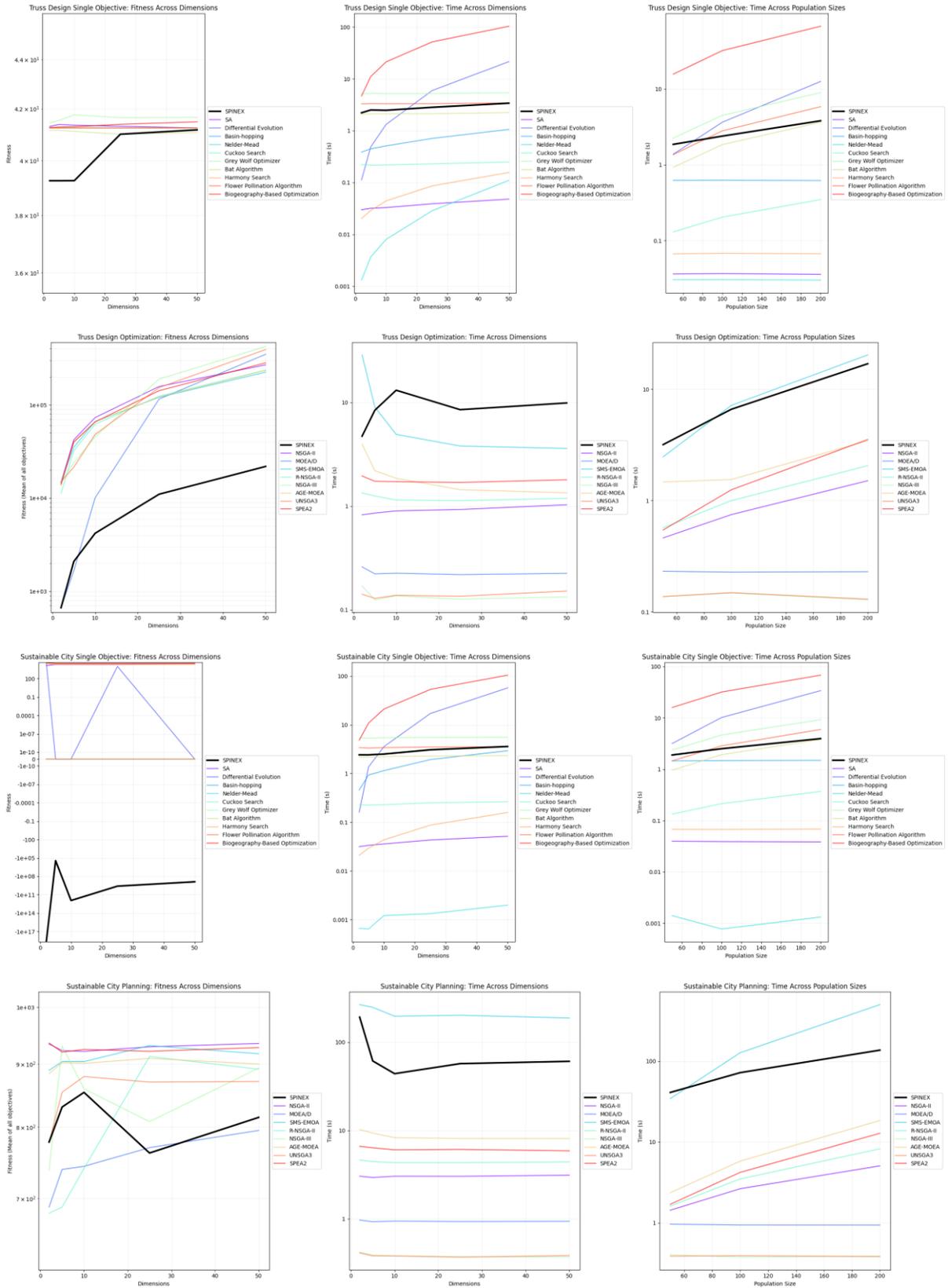

Fig. 4 Granular analysis across two unique scenarios

*4.3 Performance profile analysis*

A performance profile analysis can be used to visualize and compare the performance of multiple algorithms across a range of problems. This analysis uses the performance profiles for each algorithm and the corresponding τ values. These values represent performance ratios, which are normalized comparisons across different problem instances. The horizontal axis represents the τ, and the vertical axis represents the probability that the performance ratio of an algorithm on a given problem is within a factor τ of the best ratio. In a way, this analysis shows the best-performing algorithm (the one with the highest probability at τ = 1), the robustness of algorithms (how quickly the profiles rise), and the overall efficiency across a range of problems (the height of the profile as τ increases). Figure 5 shows the results of this analysis and notes the potential of SPINEX against other algorithms.

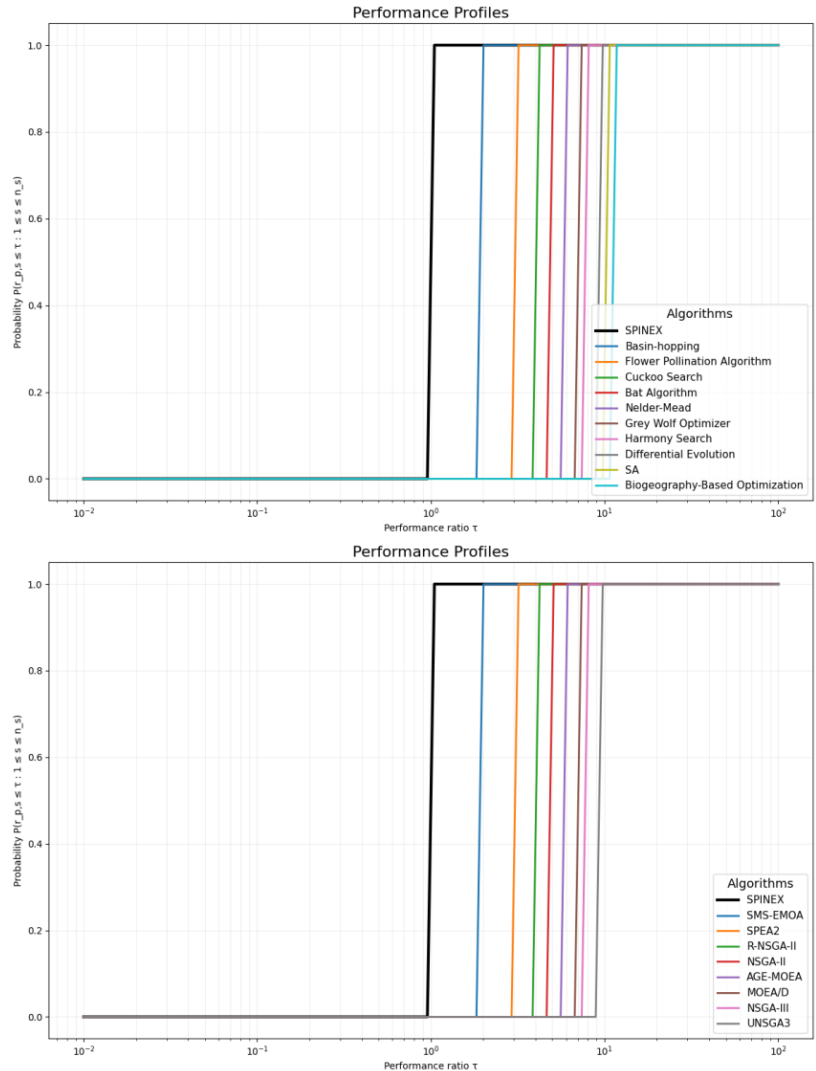

Fig. 5 Performance profile analysis (top: single objective scenario, bottom: multi- and many-objective scenarios)

*4.4 Explainability analysis*

Figure 6 presents the explainability capabilities of SPINEX. This figure shows the Neighbor Influence vs Diversity plot, which views the solution population's distribution in terms of two critical metrics: neighbor influence and neighbor diversity. This scatter plot, where each point represents a solution in the final population, reveals the complex relationship between a solution's impact on its neighbors and its distinctiveness within the population. The color gradient is mapped to fitness values such that it allows for the identification of high-performing solutions within the influence-diversity space. The plot highlights the global best solution and top performers for single objective optimization to provide context for their relative positions in this metric space. In multi-objective scenarios, it emphasizes the Pareto front solutions, offering insights into the trade-offs between objectives and their relationship to influence and diversity. The quadrant demarcations further segment the solution space as a means to facilitate the identification of solutions with specific influence-diversity profiles, such as those with high influence but low diversity, which may indicate local optima or promising search directions.

Then, the Fitness Landscape visualization presents a three-dimensional representation of the solution space for single-objective optimization problems. This plot projects the multi-dimensional solution space onto two principal components and maps fitness to the z-axis to better visualize the problem's topography. The surface plot, interpolated from discrete solution points, illustrates the continuity and gradients within the fitness landscape. Hence, the scatter of actual solutions overlaid on this surface allows for the assessment of population distribution and clustering tendencies. This visualization is particularly valuable for identifying challenging problem characteristics such as multi-modality, deceptive local optima, or vast neutral regions that may influence algorithm performance.

The Solution Similarity Matrix, presented as a heatmap, encapsulates the pairwise similarities between all solutions in the population. This matrix contains color intensity to indicate the degree of similarity can help reveal the overall diversity of the population and the presence of solution clusters. High-similarity regions along the diagonal suggest the formation of niches or convergence to specific areas of the solution space. Off-diagonal patterns can indicate the presence of distinct solution families or the algorithm's ability to maintain diverse solutions. The granularity of this visualization allows for detailed analysis of population structure and can inform decisions about diversity preservation mechanisms or niching strategies.

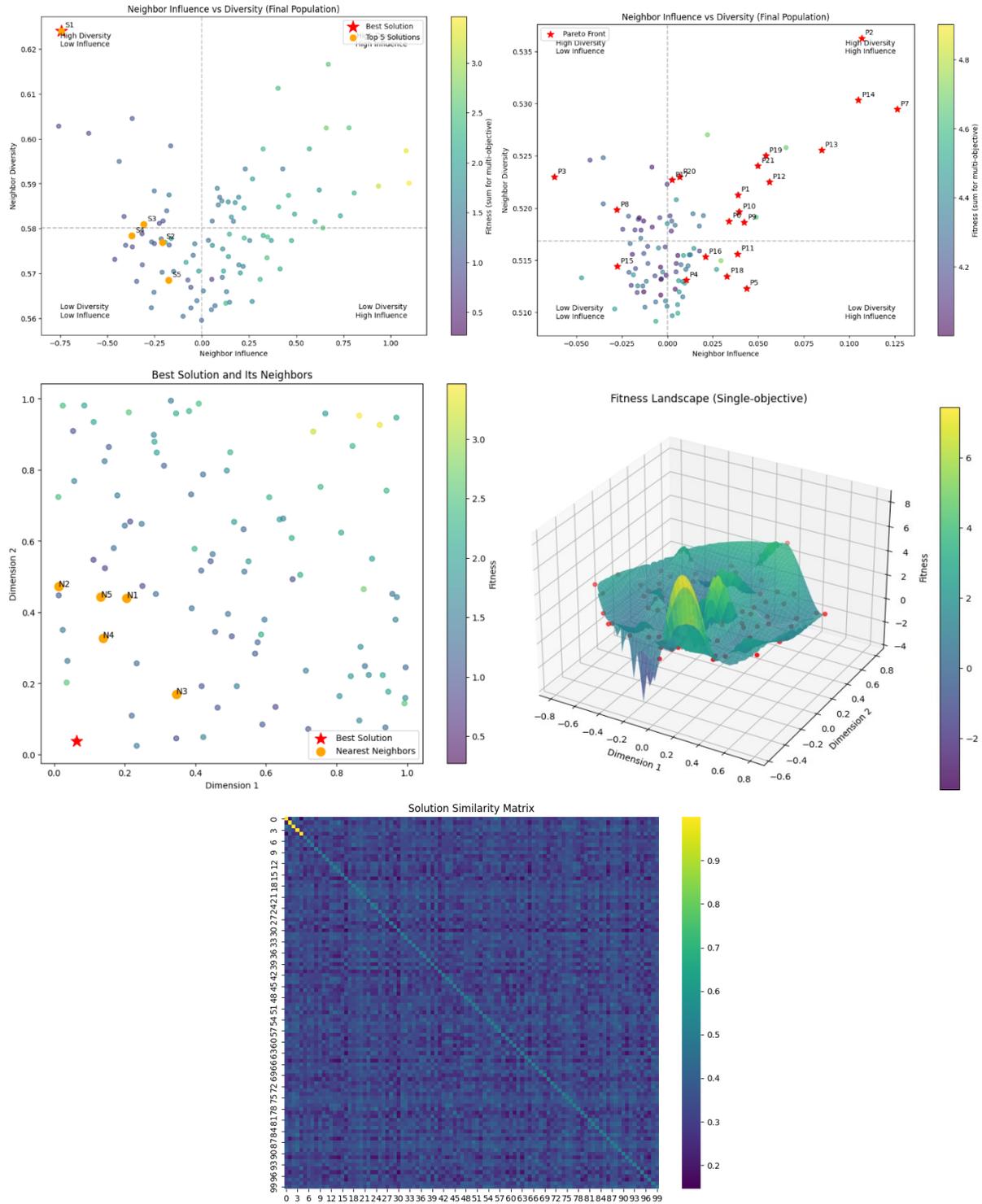

Fig. 6 Example of explainability capability of SPINEX

## 5.0 Conclusions

Our findings from conducting 55 experiments with various scenarios and complexities indicate that SPINEX-Optimization (Similarity-based Predictions with Explainable Neighbors Exploration) consistently ranks among the top-3 best-performing optimization algorithms for single, multiple, and many objective scenarios. This showcases the merit of incorporating the concept of similarity in the optimization domain. In addition, the algorithm's explainability features, Pareto efficiency, and good scalability are additional features that showcase the strength and potential of SPINEX.

**Data Availability**
Some or all data, models, or code that support the findings of this study are available from the corresponding author upon reasonable request.

SPINEX can be accessed from [**to be added**].

**Conflict of Interest**
The authors declare no conflict of interest.